\RequirePackage{fix-cm}
\documentclass{svjour3}                     
\smartqed  
\usepackage{graphicx}
\usepackage{lineno}

\begin{document}

\title{
A survey of exoplanet phase curves with Ariel
}


\author{Benjamin Charnay  \and Jo\~ao M.\ Mendon\c ca \and Laura Kreidberg  \and Nicolas B.\ Cowan  \and Jake Taylor  \and Taylor J.\ Bell \and Olivier Demangeon \and Billy Edwards \and Carole A. Haswell  \and Giuseppe Morello \and
Lorenzo V. Mugnai  \and Enzo Pascale  \and Giovanna Tinetti \and Pascal Tremblin   \and Robert T.\ Zellem}


\institute{B. Charnay \at
              LESIA, Observatoire de Paris, Universit\'e PSL, CNRS, Sorbonne Universit\'e, Universit\'e de Paris, 5 place Jules Janssen, 92195 Meudon, France. \\
              \email{benjamin.charnay@obspm.fr}           
           \and
           J. M. Mendon\c ca \at
           National Space Institute, Technical University of Denmark, Denmark
           \and 
           L. Kreidberg  \at
           Max Planck Institute for Astronomy, Heidelberg, Germany
           \and 
           N. B. Cowan  \at
           Department  of  Earth  \&  Planetary  Sciences,  Department  of  Physics,  McGill  University, Montr\'eal, Canada
           \and 
           J. Taylor  \at
           Department of Physics (Atmospheric, Oceanic and Planetary Physics), University of Oxford, UK
           \and 
           T. J. Bell \at
           Department of Physics, McGill University, 3600 rue University, Montr\'eal, QC H3A 2T8, Canada
           \and 
           O. Demangeon \at
           Instituto de Astrofísica e Ciências do Espaço, Universidade do Porto, Porto, Portugal
           \and 
           B. Edwards \at
           Department of Physics and Astronomy, University College London, London, UK
           \and 
           Carole A. Haswell \at
           School of Physical Sciences, The Open University, Walton Hall, Milton Keynes, MK17 8TT, UK 
           \and 
           G. Morello \at
           AIM, CEA, CNRS, Universit\'e Paris-Saclay, Universit\'e Paris Diderot, Sorbonne Paris Cit\'e, Gif-sur-Yvette, France
           \and
           L. V. Mugnai  \at
           Dipartimento di Fisica, La Sapienza Università di Roma, P.le A. Moro 2, 00185 Roma, Italy
           \and 
           E. Pascale  \at
           Dipartimento di Fisica, La Sapienza Università di Roma, P.le A. Moro 2, 00185 Roma, Italy
           \and 
           G. Tinetti \at
           Department of Physics and Astronomy, University College London, London, UK
           \and 
           P. Tremblin  \at 
           Université Paris-Saclay, UVSQ, CNRS, CEA, Maison de la Simulation, 91191, Gif-sur-Yvette, France
           \and 
           R. Zellem \at
           Jet Propulsion Laboratory, California Institute of Technology, Pasadena, USA
}

\date{Received: date / Accepted: date}

\maketitle

\begin{abstract}
The ESA-Ariel mission will include a tier dedicated to exoplanet phase curves corresponding to $\sim$10$\%$ of the science time.
We present here the current observing strategy for studying exoplanet phase curves with Ariel. We define science questions, requirements and a list of potential targets. We also estimate the precision of phase curve reconstruction and atmospheric retrieval using simulated phase curves.
Based on this work, we found that full-orbit phase variations for 35-40 exoplanets could be observed during the 3.5-yr mission. This statistical sample would provide key constraints on atmospheric dynamics, composition, thermal structure and clouds of warm exoplanets, complementary to the scientific yield from spectroscopic transits/eclipses measurements.

\keywords{Exoplanets \and Ariel space mission \and Atmospheres \and Phase curves}
\end{abstract}

\section{Introduction}
\label{intro}

The ESA-Ariel mission \cite{tinetti18} due for launch in 2029, will conduct a survey of $\sim$1000 exoplanets by transit and/or secondary eclipse spectroscopy. Ariel's main goals are to investigate:
\begin{enumerate}
\item the physical processes shaping planetary atmospheres
\item the chemical diversity of exoplanets
\item the formation and evolution of planetary systems
\end{enumerate}

Ariel is a 1-m telescope with simultaneous coverage from 0.5 to 7.8 $\mu$m. Data will be collected by 3 instruments: FGS (3 photometric channels: VIS-Phot, FGS-1 and FGS-2, from 0.5 to 1.1 $\mu$m), NIRSpec (1.1-1.95 $\mu$m and R$\sim$10) and AIRS (two channels: AIRS-CH0 for 1.95-3.9 $\mu$m with R=100-200; AIRS-CH1 for 3.9-7.8 $\mu$m with R=30-60).
Ariel's observational strategy is based on a 4-Tier approach. Tier 1 (Reconnaissance survey) will consist of a general transit survey at low spectral resolution (photometric bands) of $\sim$1000 exoplanets to detect primary atmospheres and clouds. Tier 2 (Deep survey) aims at characterizing the chemical composition, clouds and thermal structure of $\sim$500 planets (transits and eclipses at R$\sim$100 for AIRS-CH0 and R$\sim$30 for AIRS-CH1). Tier 3 (Benchmark planets) will consist of a detailed analysis of $\sim$50 planets amongst the best targets (transits and eclipses at R$\sim$200 for AIRS-CH0 and R$\sim$60 for AIRS-CH1). Finally, Tier 4 will correspond mostly of phase curve observations. Over Ariel's 3.5-yr mission, roughly 10$\%$ of the science time will be dedicated to phase curves. 

The principle of phase curve observations is to measure the radiation reflected or emitted by an exoplanet throughout its orbit around its star. Planets are inherently three-dimensional objects, and only by observing their phase variations can we obtain a complete picture of the global atmospheric chemistry and climate. Ariel has several advantages compared to other space missions for studying phase curves. Firstly, Ariel's spectral bands cover most of the thermal emission of warm/hot planets allowing accurate estimates of atmospheric heat redistribution and Bond albedo. Secondly, simultaneous observations with the different Ariel's channels may enable us to disentangle reflected light from thermal emission for favourable cases. Finally, phase curves are very time consuming so the forthcoming Jame Webb Space Telescope (JWST) will probably only observed one to three dozen exoplanet phase curves as explained in the next section. Conversely, Ariel provides a unique opportunity to perform a statistical survey of exoplanet spectroscopic phase curves.

We present here the plan defined by the Ariel Phase Curve Working Group to measure thermal phase curves of 35-40 exoplanets. We start in Section \ref{sec:2} with a brief review of phase curves to highlight their interest, lessons from past measurements and expectations for the coming decade with JWST. In Section \ref{sec:3}, we outline five science questions to be investigated with Ariel phase curves. We describe the target list and the observational strategy in Section \ref{sec:4}. We then discuss the effect of systematic error and the expected precision for atmospheric retrieval based on simulated phase curves from a 3D model (Section \ref{sec:5}). We finish with a summary and conclusions in Section \ref{sec:6}.

\section{Brief Review of Phase Curves}
\label{sec:2}

Thermal phase curves of a short-period planet constrain its atmospheric circulation—specifically the transport of energy from the day to night hemisphere of a synchronously-rotating planet (for a recent review, see \cite{parmentier2018}). The combination of thermal eclipse depths and phase amplitudes allows one to estimate a planet’s large-scale energy budget: how much incident flux is absorbed, and where it moves before being re-radiated to space. In other words, thermal phase curves enable the construction of simple Trenberth diagrams for exoplanets \cite{read2016}. Phase variations can also be inverted to produce a longitudinal map of the planet \cite{cowan2008}. There are now excellent publicly-available codes to perform these inversions numerically \cite{louden2018} and analytically \cite{luger2019}. 

Planets on eccentric orbits present different challenges and opportunities \cite{langton2008}. They exhibit eccentricity seasons and hence do not have a static temperature map and cannot be mapped via phase variations. On the other hand, the time-variable incident flux allows us to break the degeneracy between advective and radiative timescales that plagues circular phase curves \cite{cowan2011a,kataria15}. This opportunity has been leveraged for most of the nearby eccentric hot Jupiters \cite{laughlin2009,lewis2013,deWit2016,deWit2017}.

\subsection{Review of Previous Phase Curve Measurements}

Phase curves have so far been published for more than two dozen planets, and observations are in hand, but not yet published, for about two dozen more. The vast majority of these phase curves were obtained with the Spitzer Space Telescope \cite{werner2004}, primarily as part of the post-cryogenic ``warm'' mission. Multi-epoch phase curves and/or phase curves of non-transiting planets are fraught with astrophysical and detector degeneracy and hence have been less useful \cite{harrington2006,cowan2007,krick2016}. On the other hand, the out-of-eclipse baseline can constrain the phase variations of short-period planets \cite{wong2014}.

Due to its Earth-trailing orbit, Spitzer was capable of uniterrupted monitoring of an entire planetary orbit. Indeed, continuous Spitzer phase curves of transiting (and eclipsing) planets have been the most useful, starting with the continuous, half-orbit observations \cite{knutson2007}. Full-orbit continuous phase observations \cite{cowan2012} offer a look at all of the longitudes on the planet, as well as providing two eclipses with which to remove first-order detector and astrophysical noise (the in-eclipse system flux should be the same at the start and the end of the observations).   

More recently, a few near-infrared phase curves have been measured with the Hubble Space Telescope (HST) \cite{stevenson14,kreidberg2018a,arcangeli2019}. HST phase curves have two advantages: 1) the NIR coincides with the peak dayside emission of many hot Jupiters, and 2) HST enables spectroscopic phase curves.
HST has offered us a glimpse of what exoplanet characterization will be like in the era of JWST, but HST’s low Earth orbit makes it impossible to obtain continuous measurements throughout a planet’s orbit. Full-orbit phase curves are therefore stitched together, which requires a model of the telescope and detector systematics, which can lead to increased astrophysical uncertainty \cite{stevenson14,keating2017,louden2018,mendoca2018,keating2019,morello2019}.

Lastly, optical phase curves have been observed for a handful of planets with Kepler and TESS. For the hottest planets, these phase curves still primarily probe thermal emission from the planet, while for cooler planets they probe reflected starlight \cite{parmentier2016}. In cases where the temperature of the planet can be established with thermal measurements, it has been possible to measure not only the geometric albedo of an exoplanet but also to map the albedo across the dayside of the planet with an optical phase curve \cite{demory2013}. A planet’s reflected light can also be disentangled from its thermal emission with polarimetric phase measurements, but these measurements have so far proven exceedingly difficult and rare \cite{berdyugina2011,wiktorowicz2015,bott2016}.

\subsection{Trends from Past Measurements}

The past decade of exoplanet phase curve measurements has revealed a few intriguing trends and some notable exceptions. First of all, it has been observed that the day-to-night temperature contrast increases with equilibrium temperature \cite{cowan2011b,schwartz2015}, likely due to the shorter radiative relaxation timescale of hotter atmospheres \cite{showman2002}.  However, the strong temperature dependence of the radiative timescale has made it difficult to tease out the variations in wind speed \cite{perez-becker2013,komacek2016,tan2019}. 

Phase curve measurements have also uncovered trends in albedo: the Bond albedos inferred from Spitzer thermal phase measurements are significantly greater than the geometric albedos inferred from Kepler optical eclipse measurements \cite{schwartz2015}, a dichotomy known as the “Albedo Problem” \cite{crossfield2015}.  One hypothesis is that most hot Jupiters are dark at optical wavelengths but reflective in the near-infrared \cite{schwartz2015}; this can be tested with NIR spectroscopic or ---ideally--- polarimetric observations. More recently, it was shown that the Bond albedo inferred from thermal phase curves is negatively correlated with planetary mass \cite{zhang2018}.  If corroborated, this trend could be due to metallicity or surface gravity, but the details remain to be worked out.  

The nightside flux of a synchronously rotating exoplanet is a direct measure of heat transport: any power radiated from the nightside must have been conveyed there from the dayside. Two recent studies have noted that the nightside effective temperatures of hot Jupiters are remarkably uniform at $\sim$1100 K, independent of their equilibrium temperature, $T_{\rm eq}$, or dayside effective temperature, $T_{\rm day}$ \cite{keating2019,beatty2019}, possibly due to optically thick nightside clouds. The ultra-hot Jupiters, on the other hand, exhibit an increase in nightside temperature (and larger phase offsets \cite{zhang2018}), likely due to hydrogen dissociation and recombination \cite{bell2018,tan2019,mansfield2020}.

The broad trends observed among exoplanets have occasionally been bucked by individual planets. Most thermal phase curves peak before the eclipse, indicative of eastward winds, as predicted by global circulation models \cite{showman2002}. The hotspot offset of HAT-P-7b as seen in the optical with Kepler, however, seems to change direction over time \cite{armstrong2016}, while the Spitzer phase curve of CoRoT-2b exhibits a significant westward offset \cite{dang2018}. Likewise, while most hot Jupiters appear to have optical geometric albedos of a few percent \cite{heng2013}, Kepler-7b is about an order of magnitude more reflective \cite{demory2013}. Lastly, while nightside temperatures generally seem to be a weak function of equilibrium temperature, the hot Jupiter WASP-43b has been reported to have a very low nightside temperature \cite{stevenson14}, at odds with other hot Jupiters with similar equilibrium temperatures.  
If taken at face value, the above oddballs suggest that planetary age (CoRoT-2b), gravity (Kepler-7b), and orbital period (WASP-43b) significantly affect the large-scale absorption and transport of energy on hot Jupiters. Indeed, comparing phase curves of exoplanets with similar bulk properties is a promising approach to establish the fundamental parameters governing planetary climates \cite{Keating2020}.	

\subsection{What to expect from JWST}

We know of four phase curve observations planned for Early Release Science or Guaranteed Time Observations: WASP-43b (MIRI/LRS ERS and NIRSpec/BOTS GTO)\footnote{Backup ERS targets are WASP-103b and KELT-16b}, WASP-121b and TOI-193.01 (NIRISS/SOSS GTO).
It is safe to say that JWST spectroscopic phase curves will be even more informative than the handful of HST phase curves due to the much greater collecting area and spectral coverage. 

It remains to be seen how many additional phase curve measurements will be observed as part of JWST General Observer (GO) programs. One can obtain a pessimistic estimate from the Hubble Space Telescope, which, like JWST, enjoys enormous proposal pressure from a wide variety of astronomers. The first HST phase curve was observed in 2013 and published the following year \cite{stevenson14}. Since that landmark paper, Hubble phase curves have been obtained and published at a rate of about one per year. The allocation of JWST guaranteed time observations, on the other hand, bodes well for phase  curves: three planets will benefit from full orbit phase measurements as part of GTO.  So one could optimistically expect that JWST will observe complete phase curves for approximately three planets per year for the duration of the mission.  Given its expected duration of 5--10 years, we therefore estimate that between 7 and 30 full orbit phase curves will be acquired by JWST over the course of the mission. We may reasonably expect the lion's share of those to be for planets with orbital periods shorter than one day, the cutoff between small and medium proposals.  Regarding planets with longer orbital periods, it would make sense to point JWST at temperate and terrestrial planets since it is the only space telescope able to meaningfully characterize their atmospheres and climate \cite{cowan2015}.

\section{Science questions for Ariel phase curves} 
\label{sec:3}
\begin{figure} 
\begin{center} 
	\includegraphics[width=5cm]{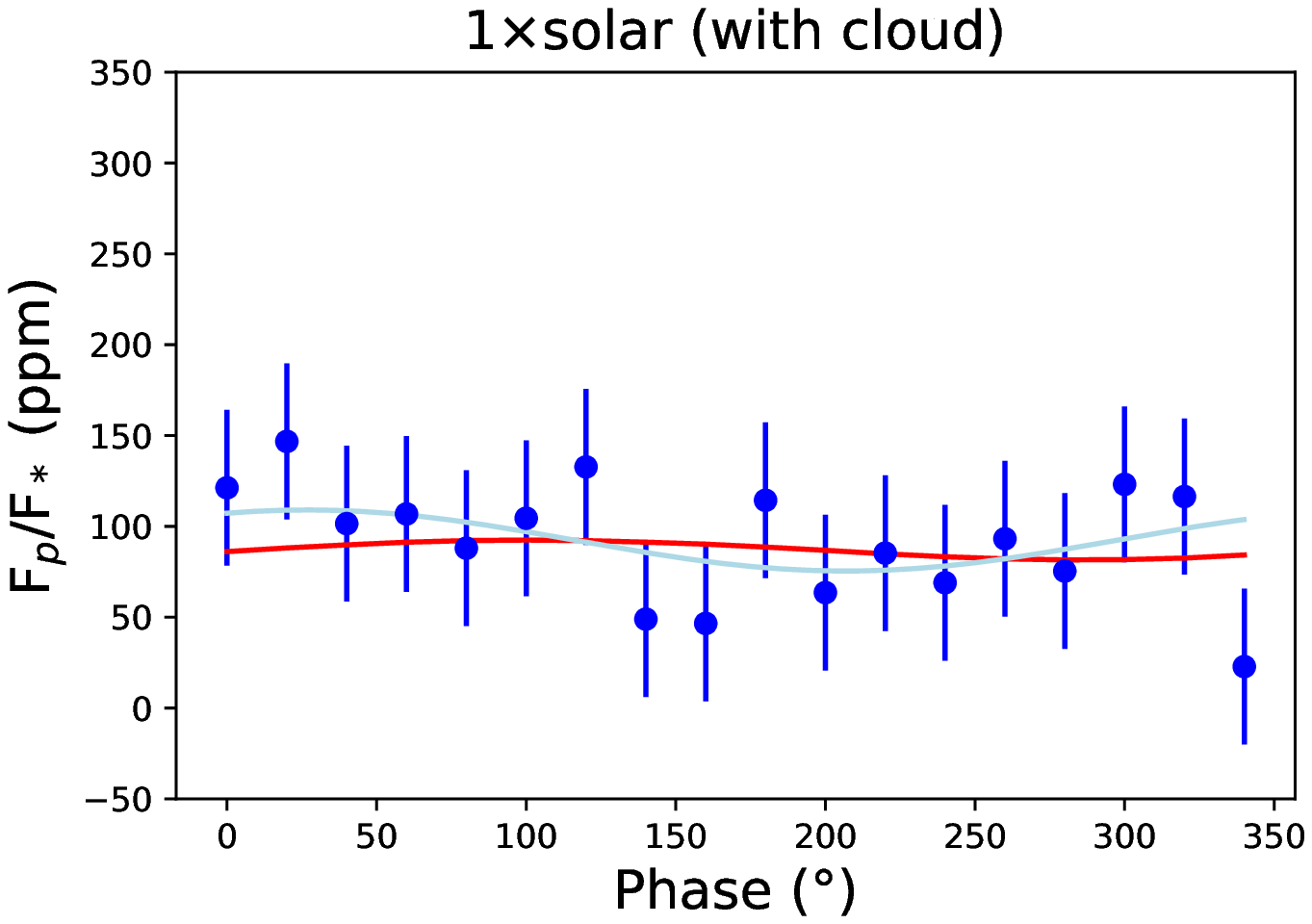}
	\includegraphics[width=5cm]{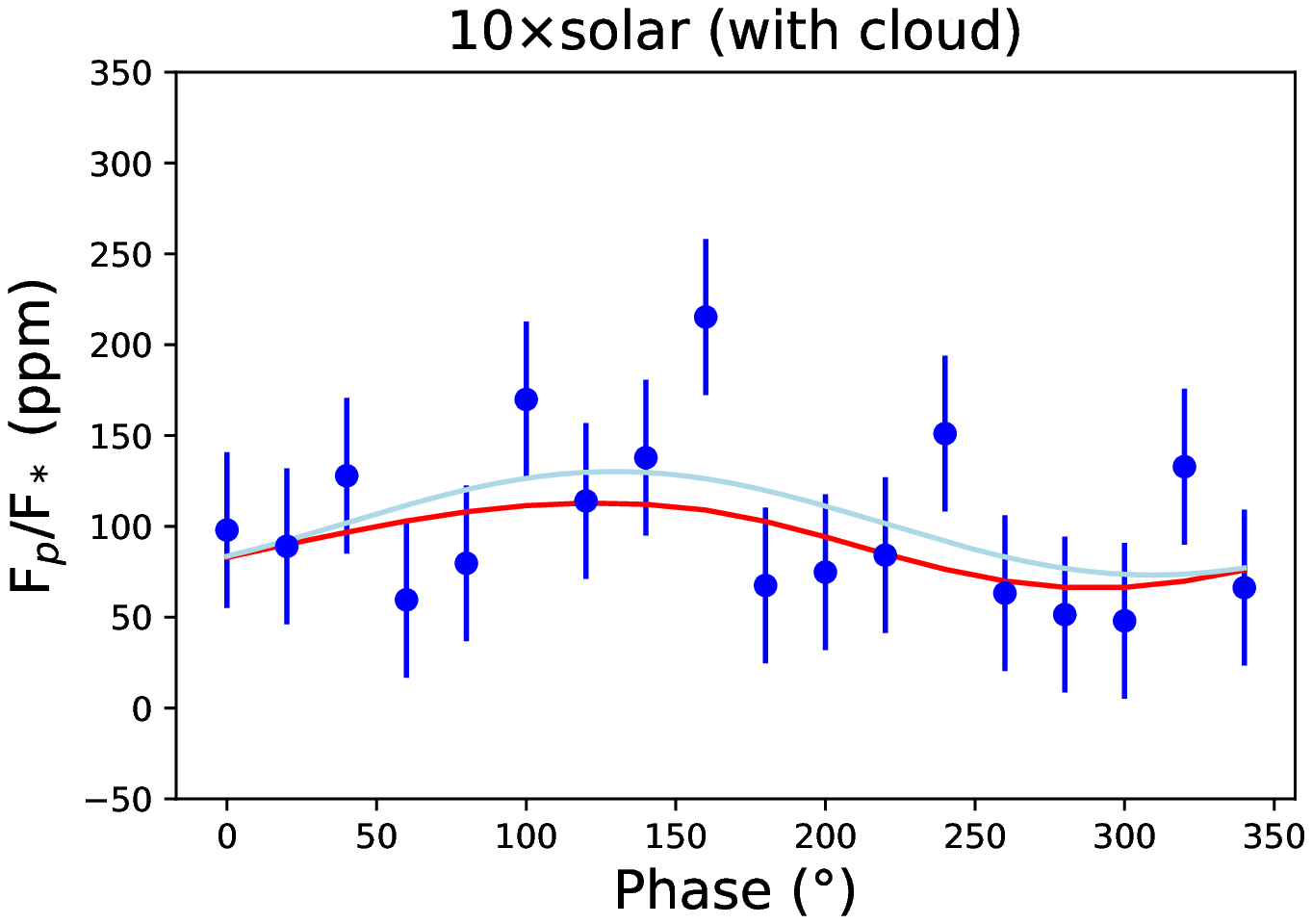}
	\includegraphics[width=5cm]{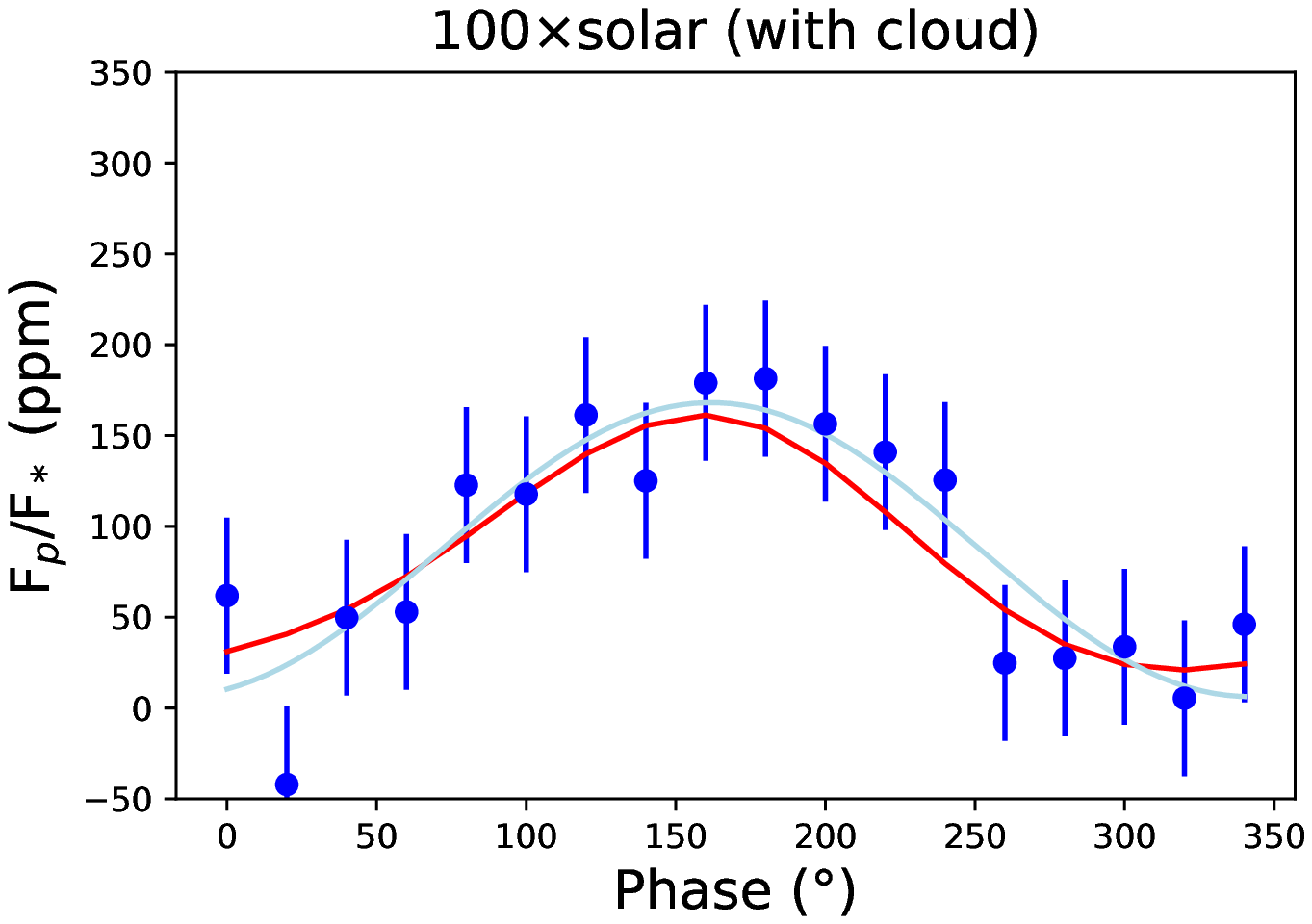}
	\includegraphics[width=5cm]{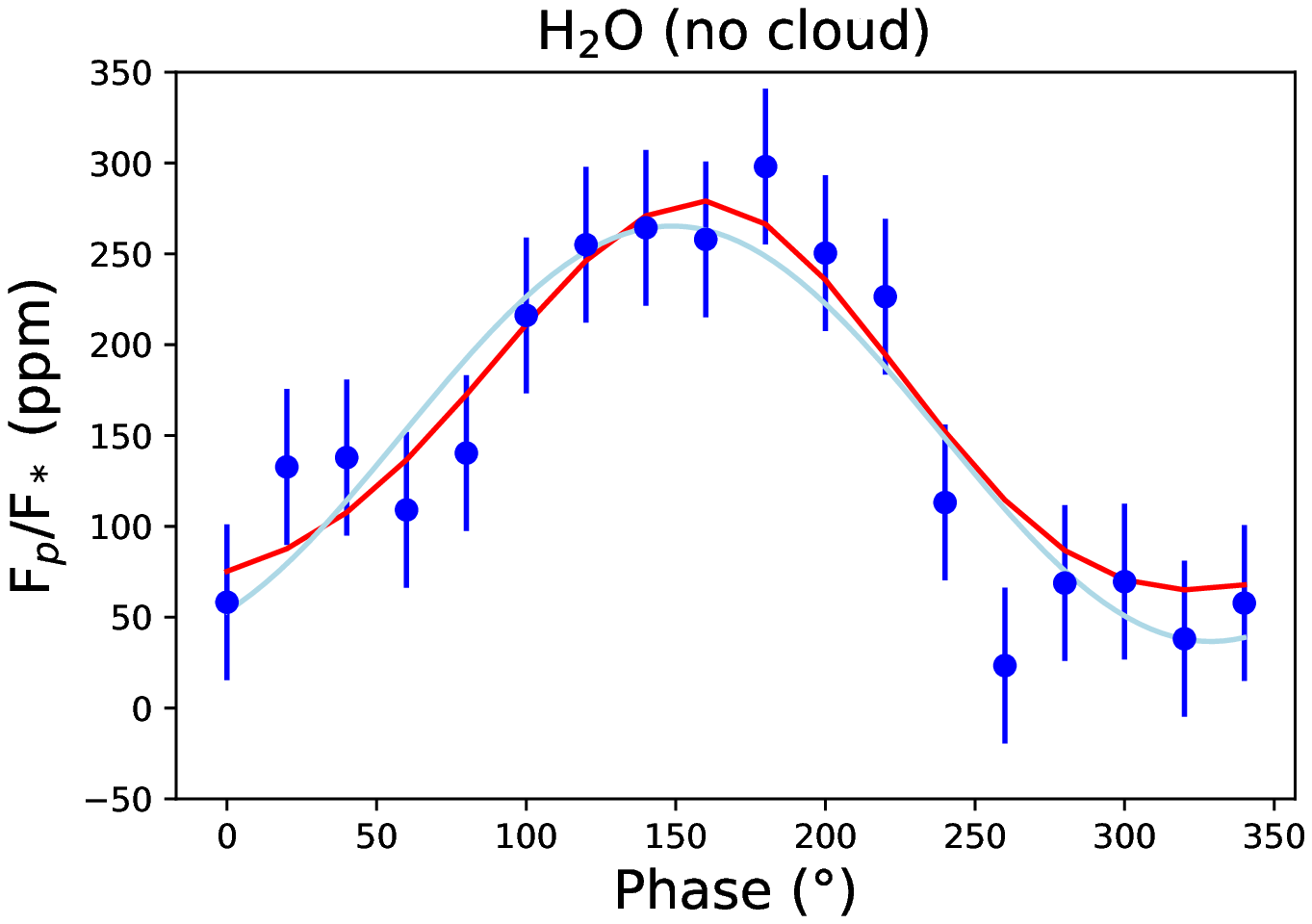}	
\end{center}  
\caption{Simulated phase curve of GJ 1214b for one orbit with AIRS-CH1 (3.9-7.8 micron) and for different atmospheric compositions. The red line is the GCM flux variations \cite{charnay15b}, blue points are simulated Ariel observations and the light blue curve is a sinusoidal fit. The corresponding SNR for the amplitude is: 0.38, 1.63, 4.9 and 7.5 for the solar metallicity, 10$\times$solar metallicity, 100$\times$solar metallicity and pure water case respectively. Without heat redistribution, the SNR would reach 11.1. The thermal phase curve is easily detected for high-metallicity cases ($\geq$100$\times$solar).}
\label{figure_1}
\end{figure}

We have defined five science questions to investigate with Ariel phase curves. They are focused on the coupling between atmospheric dynamics, chemical composition, thermal structure and clouds. Our objective is to answer these questions using a statistical sample of phase curves, covering a large parameter space. 
\\
\newline
\textbf{SQ1: Which parameters control atmospheric heat redistribution?}
\\
Phase curve measurements will allow us to determine atmospheric heat redistribution as a function of stellar irradiation, planetary radius, metallicity and eccentricity. Measuring dayside/nightside emission and phase offsets from the sub-stellar point of the phase curve peak for a large range of planetary parameters would provide information about the circulation regime, the atmospheric radiative timescales, wind speeds and the effect of nightside clouds. 
\\
\newline
\textbf{SQ2: How do atmospheric composition $\&$ thermal structure change from dayside to nightside?} 
\\
We expect significant variations in the chemical composition as a function of longitude for strongly irradiated planets due to thermochemistry and molecular dissociation. Atmospheric circulation should smooth out these variations, producing chemical disequilibrium. In addition, longitudinal/vertical temperature variations are expected, with stratospheric thermal inversion on the dayside of some hot Jupiters due to absorption of incoming stellar flux at low pressure.
\\
\newline
\textbf{SQ3: What is the atmospheric composition of low-mass planets?}
\\
A major question for Ariel is to understand the atmospheric composition or the atmospheric metallicity of low-mass planets. 3D climate models predict that the day-to-night heat redistribution decreases with atmospheric metallicity due to a reduction of the atmospheric radiative timescale \cite{menou12,charnay15b}. Consequently, low-mass planets with high atmospheric metallicity should have high-amplitude phase curves (see Fig. \ref{figure_1}). Phase curves can thus be used to measure independently of transit spectroscopy atmospheric metallicity. Finally, they can reveal the presence of an atmosphere on a rocky planet through the inferred heat redistribution (e.g. \cite{kreidberg19}).
\\
\newline
\textbf{SQ4: What is the albedo of exoplanets?}
\\
Most exoplanets are at least partially cloudy. A major questions is to understand the nature of these aerosols. Are they condensate clouds or photochemical hazes? The albedo of exoplanets is related to the optical properties and cover of clouds/hazes. Transitions in the values of the albedo are expected due to the disappearance of some clouds with temperature, as for the L-T transition for brown dwarfs \cite{parmentier2016}. Measuring the albedo would help to determine the nature of these aerosols. The wavelength-integrated albedo is also a fundamental parameter for the thermal balance of planets. 
\\
\newline
\textbf{SQ5: How do thermal structure and aerosols vary in time?}
\\
Atmospheric variability could be caused by shear instability and shocks, variation in cloud cover, or magnetic interactions \cite{fromang16,parmentier2016,rogers17}. Given that brown dwarfs with the same size, composition, and temperature as hot Jupiters have long been known to exhibit weather, it will come as no surprise if hot Jupiters are variable, too.  It would be useful to measure the magnitude and timescale of such variability. 3D cloud-free simulations suggest variations of the order of just 1$\%$ for the phase curve amplitude over timescales of a few days to a few weeks \cite{komacek20}. The presence of clouds may lead to a higher photometric variability (e.g. a few percent for the phase curve amplitude), as suggested by some 3D simulations \cite{charnay20} and brown dwarfs observations \cite{biller17}.

\begin{table}[!h] 
\begin{tabular}{|l|l|l|}
\hline 
   Science Questions &  Observations & Required precision \\ \hline
   SQ1 Atmospheric heat transport & photometric &  10$\%$ for amplitude and 5$^\circ$ for offset \\
   SQ2 Variations of composition & spectroscopic &  0.5 dex for molecular abundances\\
   SQ3 Low-mass planets & photometric &   0.5 dex for the metallicity   \\
   SQ4 Albedo & photometric &   0.1 for geometric and Bond albedo    \\
   SQ5 Time variability & photometric &   2$\%$ for amplitude and 3$^\circ$ for offset    \\     
\hline
\end{tabular}
\caption{Science questions and requirements for Ariel phase curves.}
\label{table1}
\end{table}

\begin{figure}[!h] 
\begin{center} 
	\includegraphics[width=8cm]{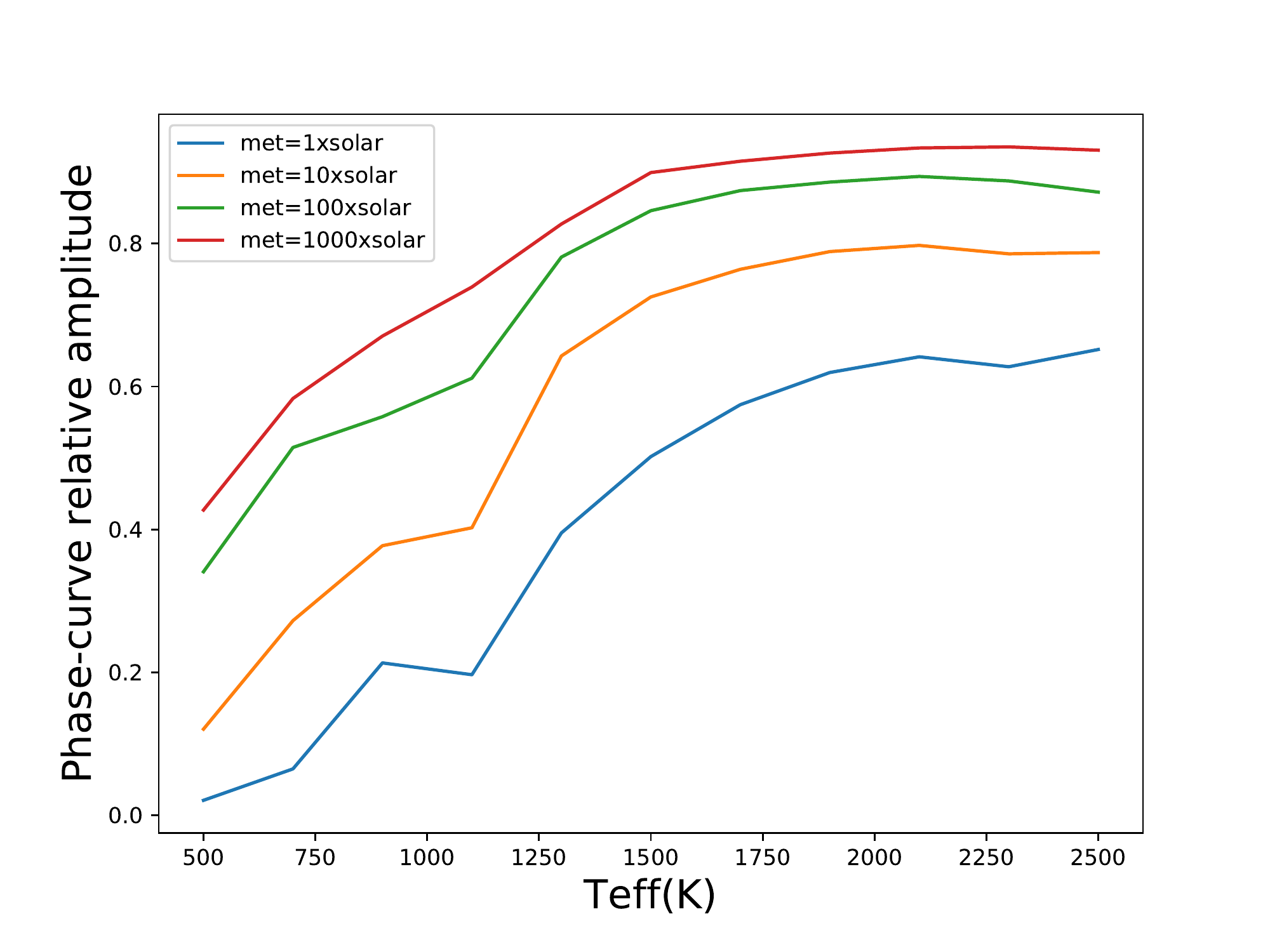}
\end{center}  
\caption{Relative phase curve amplitude in AIRS-CH1 (3.9-7.8 micron) as a function of metallicity and effective temperature from a grid of 2D ATMO models  \cite{moses20}.
Simulations performed with the radius, rotation and stellar type of GJ436b.}
\label{figure_2}
\end{figure} 
\textbf{General requirements for the science questions (SQs):}
\\
Table \ref{table1} shows the type of observations (photometric or spectroscopic phase curves) and the required precision for each science question. For most of them, the required precision is equivalent to 10$\%$ on the maximal amplitude (i.e. the case with no heat redistribution) of the phase curve.
According to Fig. \ref{figure_2}, a precision of 0.5 dex on the metallicity corresponds to 10$\%$ of precision on the maximal amplitude.
For SQ5, we chose a required precision of 2$\%$ for the amplitude (SNR$>$50).
In conclusion, the general requirement to address our science questions is to reach a signal-to-noise ratio (SNR) higher than 10 (i.e. detection at 10 sigma) for the fitting of photometric phase curves, assuming no heat redistribution and no offset. For spectroscopic phase curves, we consider that the requirement is to reach SNR$>$10 for the emission spectra at Tier 2 resolution (R$\sim$50) in each of 10 orbital phase bins.

\section{Target list and observing strategy}
\label{sec:4}
\subsection{Target list}

We constructed a list of potential targets for phase curves from the whole Ariel target list \cite{edwards19}. We divided planets into four categories:
\begin{itemize}
\item rocky planets with planetary radii lower than 1.8 R$_\oplus$
\item sub-Neptunes with planetary radii between 1.8 and 3.5 R$_\oplus$
\item Neptunes with planetary radii between 3.5 and 7 R$_\oplus$
\item Giants with planetary radii larger than 7 R$_\oplus$
\end{itemize}
We limited the sample to planets with orbital periods of less than 5 days. We classified them by the SNR of the maximal phase curve amplitude (no heat redistribution). The SNR is computed accumulating a whole number of orbits up to 10 days at Tier 1 resolution for rocky/sub-Neptunes, observing one orbit at Tier 1 resolution for Neptunes and giant planets. For the latter, we also computed the SNR for observations with a total integration time of 10$\%$ of the orbital period at Tier 2 resolution for spectroscopic phase curves.
The SNR is evaluated assuming that the phase curve is a pure sine with no offset and decomposed into observation bins of 1 hour. We first compute the SNR for a 1-hour observation at full phase using:

\begin{equation}
SNR_{\rm 1h} = \frac{F_{\rm planck}(T_{\rm p})}{F_{\rm planck}(T_{\rm s})}  \left(\frac{R_{\rm p}}{R_{\rm s}}\right)^2 \times \frac{1}{\sigma_{\rm 1h}}
\end{equation}
where $T_p$ is the planet equilibrium temperature with no heat redistribution assuming a Bond albedo of $A_B$=0.3 ($T_p<$700K) or $A_B$=0.1 ($T_p>$700K), $T_s$ is the star temperature, $R_p$ and $R_s$ are the planet and star radii. These parameters are given in the Ariel target list \cite{edwards19}. $\sigma_{1h}$ is the noise for a 1-hour observation and obtained with the Ariel Radiometric Model (\texttt{Ariel-RAD}) \cite{mugnai20}. We used the maximal value of SNR between the different Ariel's spectral bands (Tier 1 resolution).

By computing the error on the estimation of the amplitude of the phase curve assumed to be a pure sine, the SNR for a full orbit (with $P$ the orbital period in hours and $P\gg$1h) can be expressed as (see Appendix):
\begin{equation}
SNR_{\rm orbit} = 0.5\times SNR_{\rm 1h}\times \sqrt{P/2}
\end{equation}

\begin{figure}[!h] 
\begin{center} 
	\includegraphics[width=10cm]{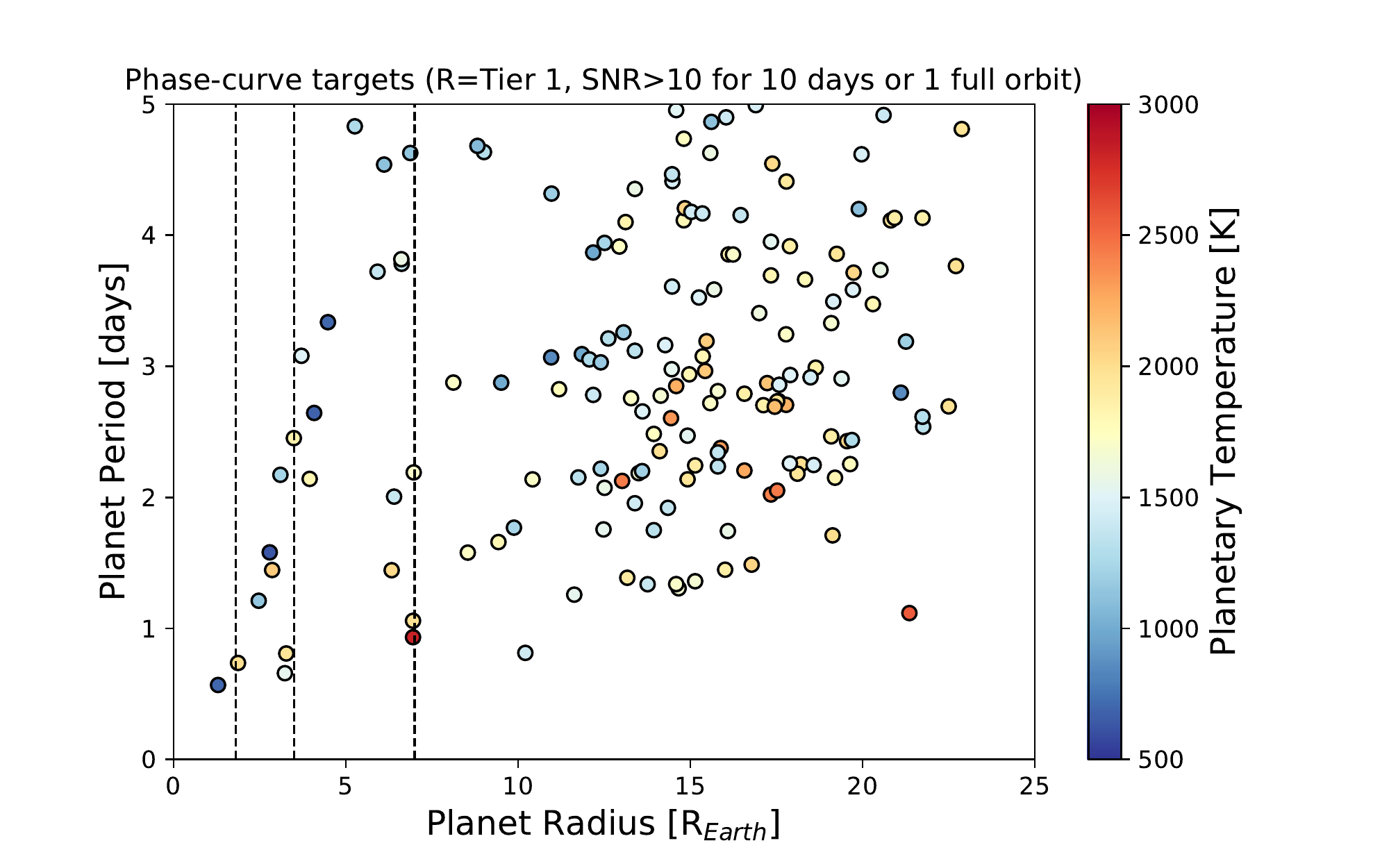}	
	\includegraphics[width=10cm]{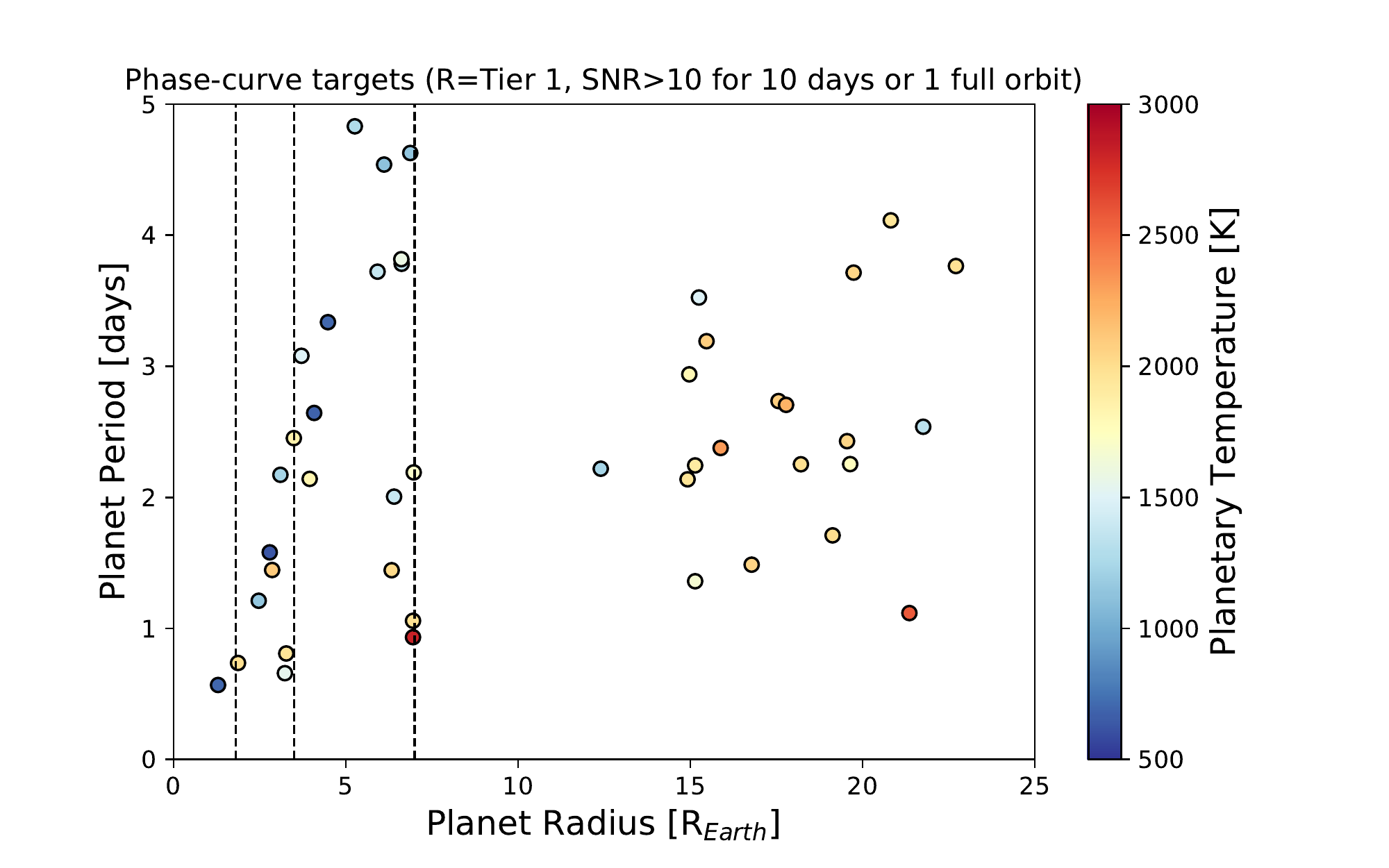}
\end{center}  
\caption{Targets for phase curves as a function of planetary radius, planet period and temperature (color bar). Vertical dashed lines represent the radius limits of the 4 categories: rocky planets, sub-Neptunes, Neptunes and giants planets. The top panel shows all targets reaching the conditions for phase curves. The bottom panel is limited to the 20 best giant planets.
}
\label{figure_3}
\end{figure}

We validated this metric with comparison to Spitzer observations of LHS3844b (see Appendix).
We limit our sample to planets reaching SNR$>$10, which corresponds to a 10-sigma detection of the phase curve, for the conditions given above for each planet category. We also limit the number of giant planets to 20. With all these conditions, we get 44 planets (see the bottom panel in Fig. \ref{figure_3}) including: 
\begin{itemize}
\item 1 rocky planet
\item 8 sub-Neptunes
\item 15 Neptunes
\item 20 giants
\end{itemize}
All planets in this sample reach SNR$>$10 for photometric (Tier 1 resolution) thermal phase curves, which is the requirement for SQ1 (heat redistribution), 3 (the composition of low-mass planets and exo-Neptunes) and 4 (albedo). All giant planets fulfill the conditions for spectroscopic phase curves (i.e. SNR$>$10 in 10$\%$ of the orbital period at Tier 2 resolution, SQ2) and for variability (SQ5). Six Neptunes reach SNR$>$5 for spectroscopic phase curves.

This target list covers a large parameter space with planetary mass from 0.01 to 4 M$_{\rm Jup}$, equilibrium temperature from 600 to 2200 K, and expected atmospheric metallicity from 1 to 300$\times$solar.
15 planets from our sample are already known, including 1 rocky planet, 2 sub-Neptunes, 2 Neptunes, 10 giant planets (see Table 2). We found that a total of 136 giant planets from the Ariel target list reach the conditions for spectroscopic phase curves (see the top panel Fig. \ref{figure_3}). They include 83 already known giant planets, most of them coming from the WASP, HAT-P and KELT transit surveys. This implies a lot of choice for the selection of giant planets.
Our remaining targets targets are simulated discoveries from the Transiting Exoplanet Survey Satellite (TESS). 
TESS is surveying nearby stars for transiting planets with near-complete sky-coverage \cite{ricker15}. TESS has delivered over 1000 planet candidates in its first year (Guerrero et al., submitted).  Some of these may be exceptional targets for phase curve observations, including several new hot Neptunes with short orbital periods and high equilibrium temperatures (i.e. LTT 9779b, HD 219666b, TOI 132b, and TOI 824b). TESS is currently on track to detect over 2000 planets around bright nearby stars, providing the large statistical sample needed for Ariel.                                     

\begin{table}[!h] 
\begin{tabular}{|c|c|c|c|c|c|}
\hline 
   Planet 		&  Period   & Orbits 	& SNR  	              & SNR & Thermal/Reflected \\
    			&   (days)  &  		    &  (thermal emission) &  (reflected light) &  (FGS)\\ \hline   
GJ 1214 b 		& 	1.58    &1		&  11.0 	&	0.3 & 0.0\\
K2-266b 		&	0.66    & 4		& 11.1	    & 	1.5 & 0.0\\
55 Cnc e 		&	0.74 	&4		&10.3	    &	0.3 & 0.7\\ \hline  
GJ 436b 		&	2.64    &1		&15.2	    &	0.9 & 0.0\\
GJ 3470b 		&	3.34    &1		&10.4	    &	0.6 & 0.0\\ \hline  
HD 189733b 	    & 	2.22	& 1 or 3& 205.2	&	3.0 & 0.04\\
HD 209458b 	    & 	3.52 	&1 or 3	&168.1	    &	2.3 & 0.2\\
XO-6b 		    &	3.77	&1		&107.6 	&	3.5 & 0.9\\
WASP-77Ab 	    &	1.36    &1		&106.1	    &	3.5 & 0.3\\
KELT-7b 		&	2.73	&1		& 105.3 	&	2.8  & 1.0\\
WASP-74b 	    &	2.14	&1		& 97.6  	&	3.4 & 0.8\\
XO-3b 		    &	3.19	&1		&95.1 	    &	3.5 & 1.0\\
WASP-82b 	    &	2.71    &1		&78.8	    &	3.0  & 1.3\\
WASP-14b 	    &	2.24    &1		&75.1	    &	2.3 & 0.7\\
KELT-14b 		&	1.71	&1		&71.1 	    &	3.9 & 0.8\\
\hline       

\end{tabular}
\caption{Known planets from the phase curve target list with the number of orbits required to reach SNR$>$10 (3 orbits are indicated for multi-epoch phase curves) and the SNR at Tier 1 resolution for thermal and reflected light phase curves. For thermal emission, we give the highest value of SNR between the different photometric channels, assuming a Bond albedo of $A_B$=0.3 ($T_p<$700K) or $A_B$=0.1 ($T_p>$700K). For reflected light curves, we give the SNR integrating the flux from 0.5 to 1.1 $\mu$m (the 3 FGS channels), assuming a geometric albedo equal to the Bond albedo. The last column is the ratio of the flux from thermal emission to the flux from reflected light integrated from 0.5 to 1.1 $\mu$m.
Horizontal lines separate sub-Neptunes, Neptunes and giant planets. Note that with our limit at 1.8 R$_{\rm Earth}$, 55 Cnc e is listed as a sub-Neptune but it is likely a rocky planet.}
\label{table2}
\end{table}

\subsection{Observational Strategy}

The plan is to perform continuous phase curve observations, even when multiple orbits are required. The observations will start and end with the secondary eclipse, which represents the reference for the phase curve. We will keep a margin of 7$\%$ of the period before and after so as not to miss the eclipse. This corresponds to a maximal error of 0.1 on the eccentricity, $e$ (the uncertainty on the eclipse time is $2eP/\pi$).
The study of variability (SQ5) with multi-epoch phase curves will be limited to one or two planets (e.g. HD 189733b) with 3 phase curves (i.e. 3 orbits).
The observations of all planets from our target list, including 2 planets with multi-epoch phase curves; would take $\sim$175 days of observing time, or $\sim$13$\%$ of Ariel science time. Removing 1 sub-Neptune, 5 giants and keeping only one planet with multi-epoch phase curves, the observing time goes down to 10$\%$ of Ariel science time, compatible with the current mission time dedicated to Tier 4. We can, therefore, expect that the final target list for Ariel phase curves will include 35-40 exoplanets and will represent a statistical sample for gas planets.
It is worth mentioning that the observation of phase curves also provide transits and eclipses. In the current observing plan for Ariel, a single transit or eclipse takes 2.5 times the transit duration. For our phase curve observing plan, the equivalent of 28$\%$ of the observation time is dedicated to transits or eclipses. This high fraction significantly reduces the cost of phase curves. It comes from the fact that we are targeting short period planets (P$<$ 5 days) with a mean period of $\sim$2 days.

\subsection{Phase curves of reflected light and synergy with TESS, CHEOPS and PLATO}
A difficulty for Ariel phase curves will be to distinguish reflected light from thermal emission. The last column in Table 2 shows the ratio of the thermal flux by the reflected light for FGS (0.5-1.1 $\mu$m). Reflected light dominates only for HD189733b and HD209458b as well as warm Neptunes and sub-Neptunes. Figure 4 shows the fraction of the total flux (thermal emission+reflected light) covered by the different ARIEL's instruments (FGS, NIRSpec and AIRS). Reflected light dominates for FGS only for effective temperatures lower than 1700K for a G star and 1500K for an M star. Therefore, the thermal emission will represent a non-negligible fraction of the flux measured with FGS for most of hot Jupiters, even if we limit the measurements to VIS-Phot (0.5-0.6 $\mu$m). However, the contribution of reflected light is almost negligible in NIRSpec and AIRS channels for these planets. It would be possible to extrapolate the thermal flux at short wavelength (i.g. in FGS channels) to extract the reflected light. This method has been applied to simulated Ariel eclipse observations \cite{zellem19}. According to this study, the geometric albedo could be measured with a precision of $\pm$0.02 for $\sim$10 giant planets from Tier 3.
In addition, inhomogeneous cloud and temperature distributions lead to differences in the shape of phase curves (i.e. the offset) between thermal emission and reflected light. This could also help to distinguish thermal emission from reflected light. Additional modeling work will be needed on this aspect.

Only giant planets from the phase curve target list are expected to reach a SNR sufficient to detect reflected light and to measure the geometric albedo (see Table2). For these planets and assuming that the reflected light can be distinguished from the thermal emission (as explained above), the geometric albedo would be measured with a precision of around $\pm$0.03 with one phase curve. This is comparable to the precision for albedo measurements from eclipses of Tier 3 planets \cite{zellem19}. Note that previous observations suggest a geometric albedo around 0.1 for most of warm/hot giant exoplanets \cite{heng2013,angerhausen2015}. For Neptunes and sub-Neptunes, a mean geometric albedo could be derived from the cumulative signal of several planets. 
For comparison, phase curves and eclipses from the Kepler telescope provided geometric albedo measurements with also a precision of around $\pm$0.01-0.03 \cite{heng2013,demory2013,angerhausen2015,parmentier2016}. The Kepler telescope had a collecting area similar as Ariel and cumulated more than 1200 days of phase curve observation for each target in the field of view of the primary mission. Ariel phase curves will generally be limited to one planetary orbit, giving a SNR typically 20-30 times lower than Kepler for the same target. However, Ariel will target much brigther stars than in the Kepler survey. The precision for the geometric albedo and the amplitude of phase curves could then be relatively close to that of Kepler.

CHEOPS has reached a photometric precision of 10-17 ppm in 1 hour for WASP-189, a magnitude 6.6 star \cite{lendl2020}. This would correspond to $\sim$30-50 ppm for a magnitude 9 star. CHEOPS, as a mono-target instrument like Ariel, will not be able to stare at numerous targets for an extended period of time, but will still be able to make significant contributions for a handful of golden targets. TESS stares at the same targets for at least 27 days and at best 1 years (without considering possible extensions of the mission) and can thus achieve exquisite phase curves over time. TESS is currently reaching a precision of around 85 ppm in similar conditions. In contrast, Ariel will have a precision of around 50 ppm with VISPhot and FGS for stars brighter than magnitude 11 in V. Taking into account the observation durations, we can expect relatively similar precisions for phase curves of reflected light between CHEOPS, TESS and Ariel.
PLATO will have a photometric precision of $\sim$30 ppm in 1 hour for bright solar-like stars (mag(V)$<$11).  In addition, the PLATO fields (around 11$\%$ of the whole sky for 2 PLATO fields of 49$^\circ$, which is the current plan) will be observed for 1-3 years, cumulating a few hundred orbits for short-period planets around bright stars. PLATO phase curves of reflected light should achieve a much higher precision than the other space telescopes mentioned above. Since there will be many giant planets (136 with the current list) reaching the requirements for Ariel phase curves, we plan to choose those which will benefit from a precise visible phase curve observed by TESS, CHEOPS or PLATO. This would be a strong synergy, combining visible and infrared phase curves from these telescopes, together with accurate stellar characterization.

\begin{figure}[!h] 
\begin{center} 
	\includegraphics[width=6cm]{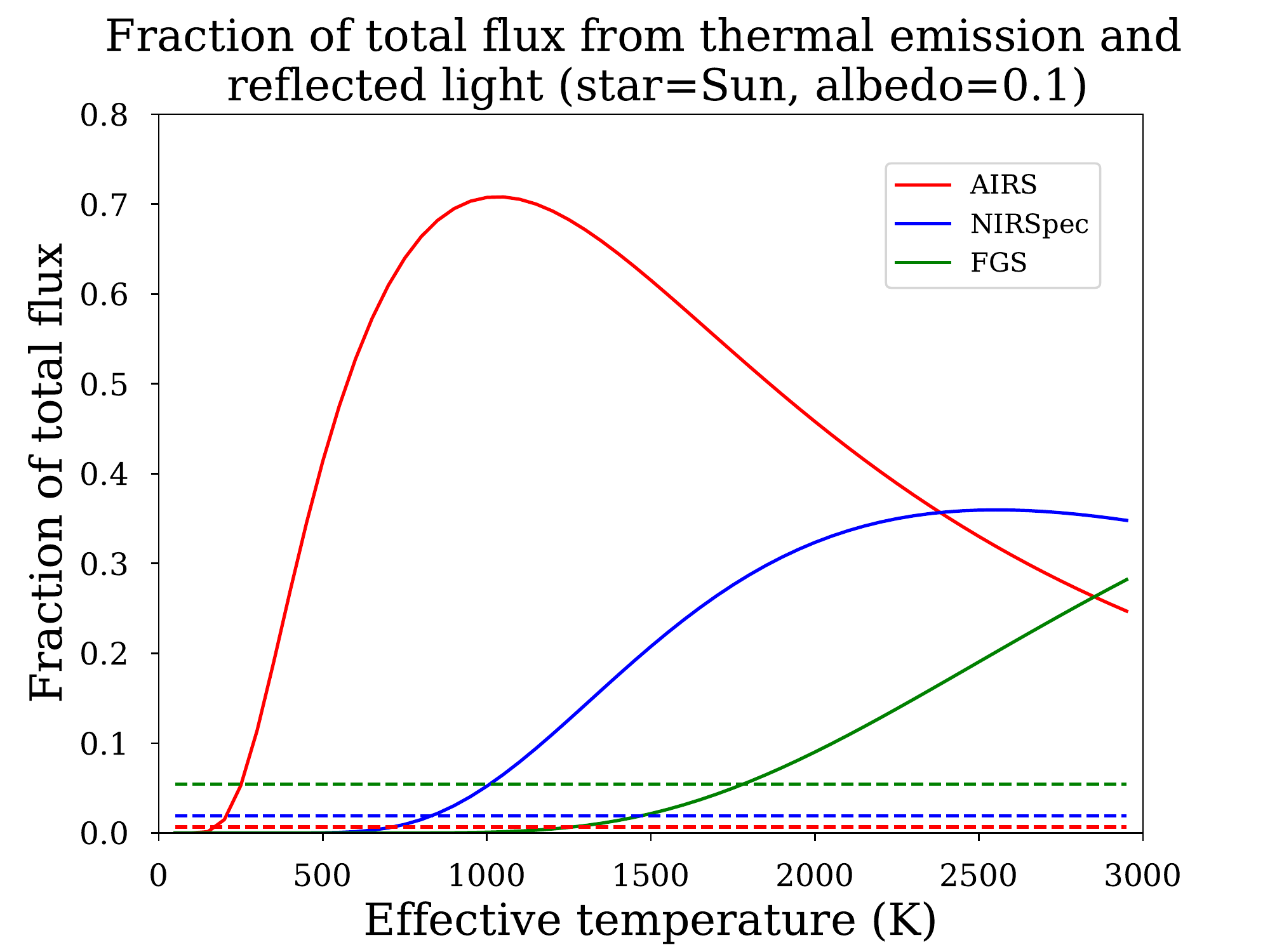}	
	\includegraphics[width=6cm]{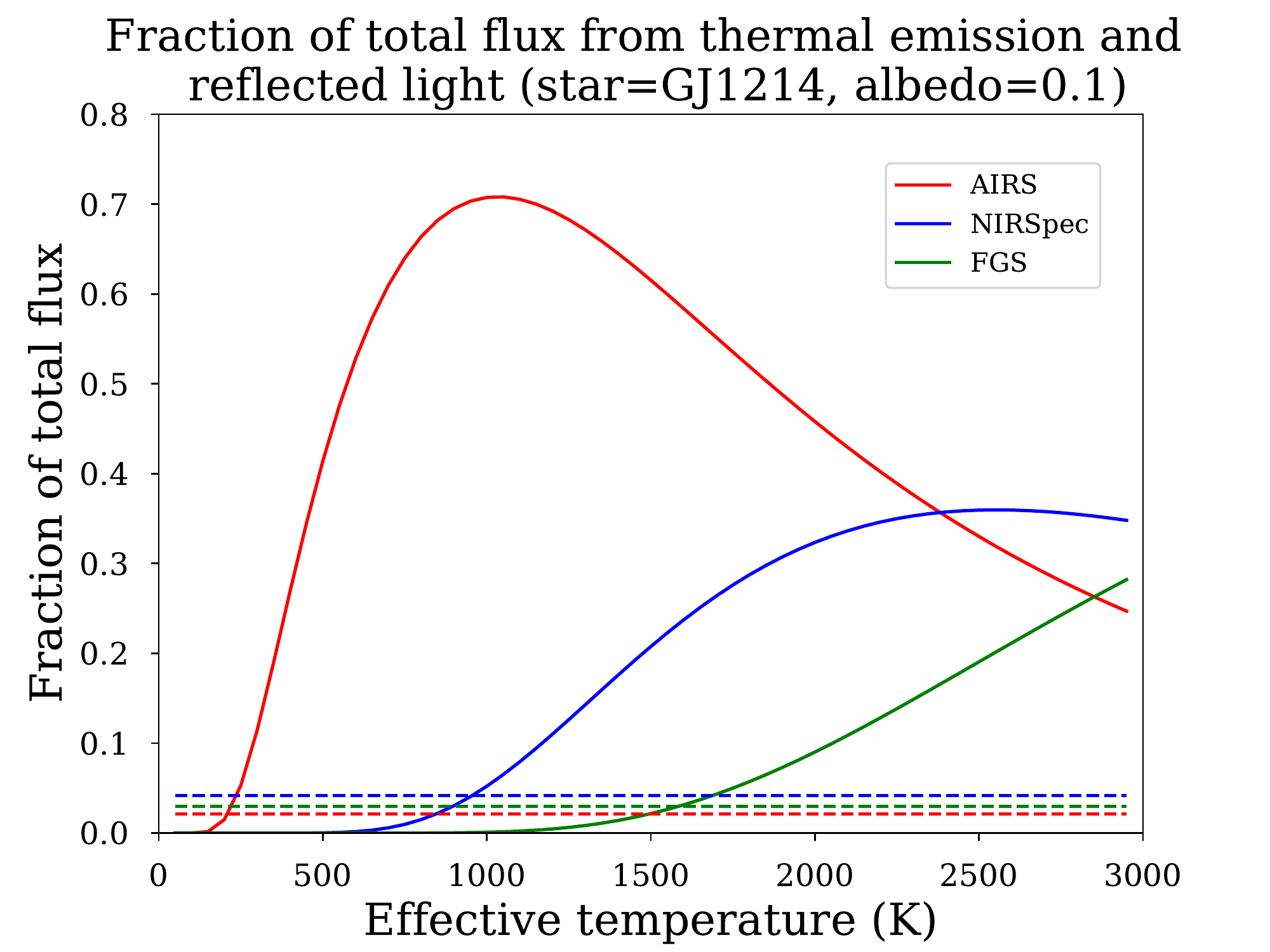}
\end{center}  
\caption{Fraction of the total flux (thermal emission+reflected light) covered by the different ARIEL's instruments (FGS, NIRSpec and AIRS). Solid lines show the thermal emission and dashed lines show the reflected flux as a function of the planetary effective temperature. The calculation is done assuming a Bond albedo and a geometric albedo of 0.1, for a Sun-like star (left panel) and for an M star as GJ1214b (right panel).
}
\label{figure_4}
\end{figure}

\section{Simulations}
\label{sec:5}
We used theoretical models to evaluate Ariel's potential performance for phase curve observations of a hot Jupiter planet. We test our theoretical tools on the planet WASP-43b and assume that the atmosphere is either cloud-free or covered by clouds on the night side. We simulated Ariel phase curves based on 3D atmospheric model results and used simulated spectra to explore our retrieval framework on Ariel observations.

\subsection{Phase curves calculated from a global circulation model}

Global Circulation Models (GCMs) are powerful tools that self-consistently simulate 3D atmospheric temperature distributions and here can predict thermal phase curves. Using a 3D GCM, \texttt{THOR}, we have simulated the phase curves of WASP-43b \cite{hellier11} with a clear atmosphere. WASP-43b has twice the mass of Jupiter, orbits its parent star in 19.2 hours with a semi-major axis of 0.0153 AU \cite{gillon12} and an estimated equilibrium temperature of 1440 K \cite{blecic14}. The parent star is a K7 star of 0.73 M$_{\rm Sun}$. 

\texttt{THOR} is a flexible GCM that can calculate phase curves and is based on a dynamical core that solves the non-hydrostatic compressible Euler equations on an icosahedral grid \cite{mendonca16}. To represent the radiative processes in the GCM simulations, we used a two-band formulation calibrated to reproduce the results from more complex codes on WASP-43b \cite{mendonca18a}. 

The multi-wavelength phase curves are obtained from post-processing the 3D simulated atmosphere with a more sophisticated radiative transfer model \cite{mendonca18a}. The spectra include cross-sections of the main absorbers in the infrared - EXOMOL: H$_2$O \cite{barber06}, CH$_4$ \cite{yurchenko14}, NH$_3$ \cite{yurchenko11}, HCN \cite{harris06}, H$_2$S \cite{azzam16}; HITEMP \cite{rothman10}: CO$_2$, CO; HITRAN \cite{rothman13}: C$_2$H$_2$. The Na and K resonance lines are also added \cite{draine11} and H$_2$-H$_2$ H$_2$-He CIA \cite{richard12}. We assumed that the atmosphere has solar abundances. As shown in \cite{kataria15}, the HST-WFC3 spectrum \cite{stevenson14} of the dayside in WASP-43b is compatible with a clear atmosphere with solar abundances. Our \texttt{THOR} simulation includes a chemical relaxation method fully coupled with the 3D dynamics for 4 different chemical species \cite{mendonca18b}: H$_2$O, CH$_4$, CO and CO$_2$. This implementation allows us to study chemical disequilibrium in the atmosphere due to atmospheric transport, which we use later to explore the potential of our retrieval framework to find departures from chemical equilibrium in future Ariel observations of WASP-43b. Other molecules in the atmosphere are assumed to be in chemical equilibrium, and their concentrations are calculated with the \texttt{FastChem} model \cite{stock18}.  The stellar spectrum was obtained from the \texttt{PHOENIX} model (\cite{allard95}, \cite{husser13}). 

\begin{figure}[!h] 
\begin{center} 
	\includegraphics[width=5.7cm]{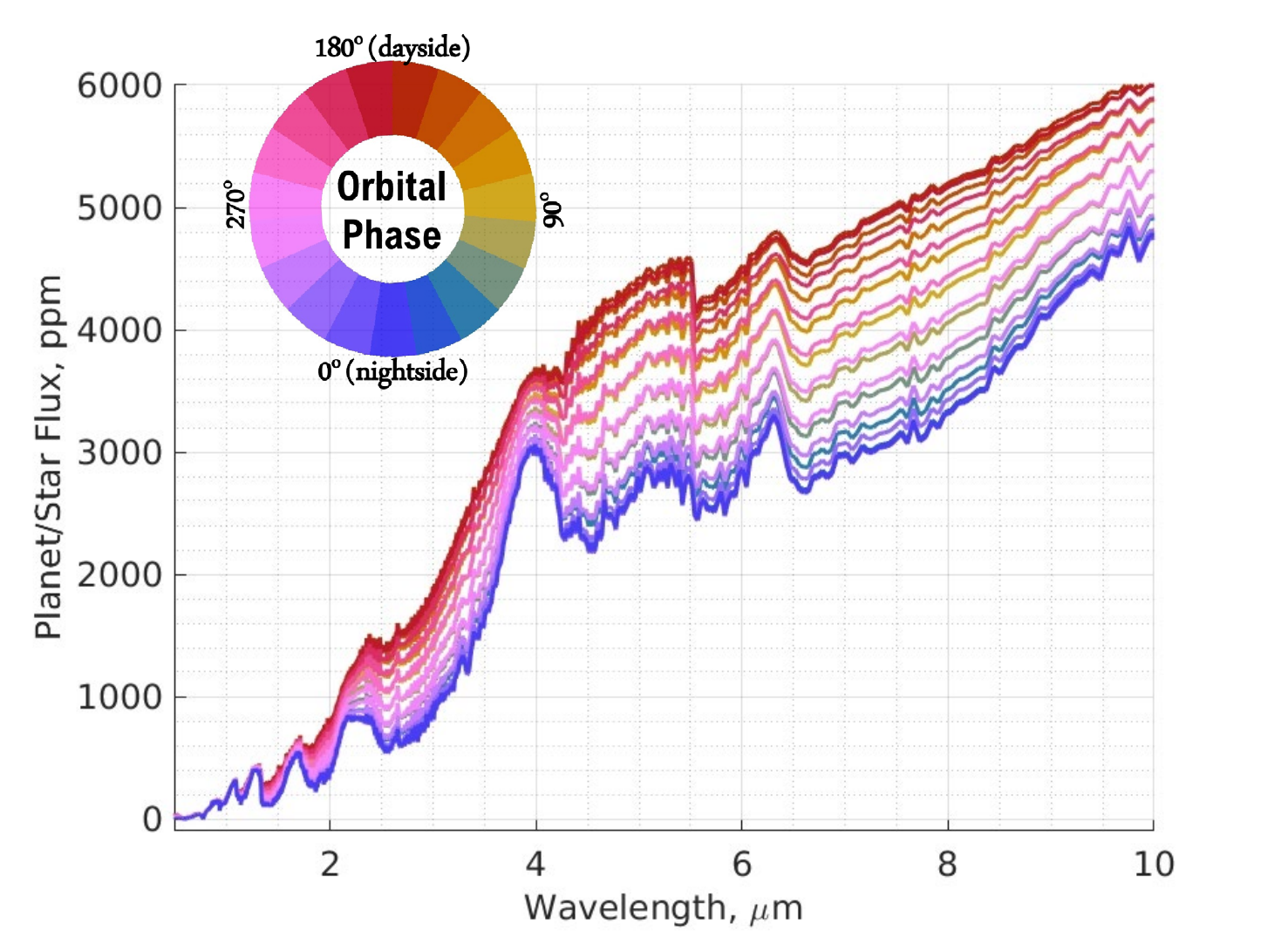}	
	\includegraphics[width=5.7cm]{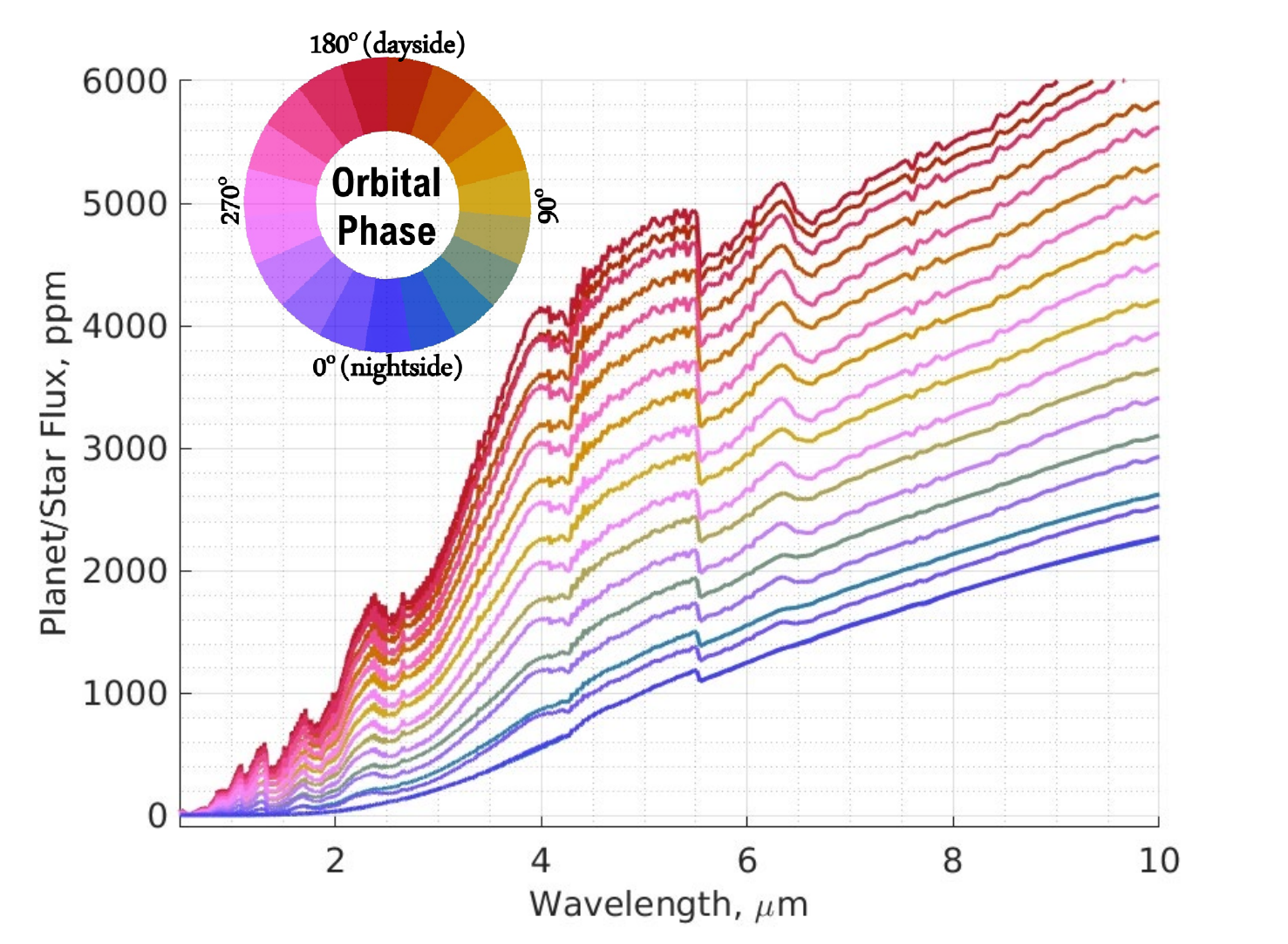}
\end{center}  
\caption{Emission plus reflected light at different orbital phases obtained with \texttt{THOR}. The circular color map shows the color for different orbital phases. The primary transit occurs at orbital phase 0$^\circ$ (blue) and the secondary eclipse at 180$^\circ$ (red). The two panels show the phase curve values for a simulation with clear sky (left panel) and a cloudy night side (right panel).}
\label{figure_model1}
\end{figure}

The disc averaged planet spectrum is calculated at each orbital phase by projecting the outgoing intensity for each geographical location of the observed hemisphere. In Fig. \ref{figure_model1}, we show the theoretical phase curves simulated with \texttt{THOR}. The different lines show the planet spectra for each orbital phase: the red lines represent the hotter dayside of the atmosphere and blue the colder nightside. The left panel in Fig. \ref{figure_model1} corresponds to the WASP-43b simulation with a clear atmosphere. We also show the impact of a grey cloud deck on the night side of the planet in the right panel of Fig. \ref{figure_model1}. The cloud distribution and optical properties are parameterized (see \cite{mendonca18a} for further details). Clouds block most of the radiative flux from the deep hot layers in the planet, which reduces the observed infrared flux for orbital phases near the primary transit \cite{mendonca18a} and reduces the spectral features.
Therefore, clouds can limit atmospheric retrieval for spectroscopic phase curves. In contrast, clouds generally make the detection of photometric phase curves easier. As shown for WASP-43b (see Fig. 4), nightside clouds tend to increase the amplitude of thermal phase curves (see \cite{parmentier2021}. In addition, dayside clouds can increase the amplitude of phase curves of reflected light and also of thermal phase curves for absorbing clouds. In this last case, the effect of clouds may remain limited compared to that of the atmospheric metallicity (see Fig. 1 and \cite{charnay15b}).

\subsection{Effect of systematics and retrieval of photometric phase curves}

Phase curve observations are a technically challenging measurement, requiring stable instrument performance spanning hours to days. The largest amplitude phase curves observed to date have peak-to-trough variation of order 0.1$\%$, but instrument systematic noise can be orders of magnitude larger than the expected signal (e.g. \cite{knutson07}). 

We carried out several simulations to evaluate Ariel's potential performance for phase curve observations. Ariel is designed for greater stability than either Spitzer or HST which have been the workhorse facilities for phase curve observations so far. One caveat is that Ariel is required to meet this stability requirement for timescales of just 10 hours, whereas nearly all phase curves have longer duration than that (see \cite{tinetti18} and Ariel Yellow Book). A work-around is to observe the phase curve in multiple overlapping pieces. To assess the feasibility of this strategy, we simulated several test light curves with varying assumptions about instrument performance.

We evaluated three potential cases: 
\begin{enumerate}
\item An optimistic case, with a continuous stare over the full phase curve with no instrument systematic noise.
\item A realistic case, with the phase curve split into three visits. Each visit is separated by a constant offset drawn from a normal distribution with mean 0 and standard deviation 10$^{-3}$. We considered two sub-cases: one with an overlap between visits of 10$\%$ in orbital phase, and the other with a 25$\%$ overlap.
\item A pessimistic case, same as case (2) above except with an additional linear trend for each visit, also drawn from a normal distribution with mean 0 and standard deviation 10$^{-3}$.
\end{enumerate}

The amplitude of the systematic noise was inspired by long-duration Spitzer and Hubble observations, which show long-term systematic trends and offsets that are still poorly explained (e.g. \cite{kreidberg18}). Spitzer and Hubble's data also show additional noise from intrapixel sensitivity variations and charge trapping (\cite{long15}, \cite{ingalls16}, \cite{zhou17}), but Ariel observations are less likely to be affected by these noise sources thanks to the stability requirements and uninterrupted pointing.

\begin{figure}[!h] 
\begin{center} 
	\includegraphics[width=6cm]{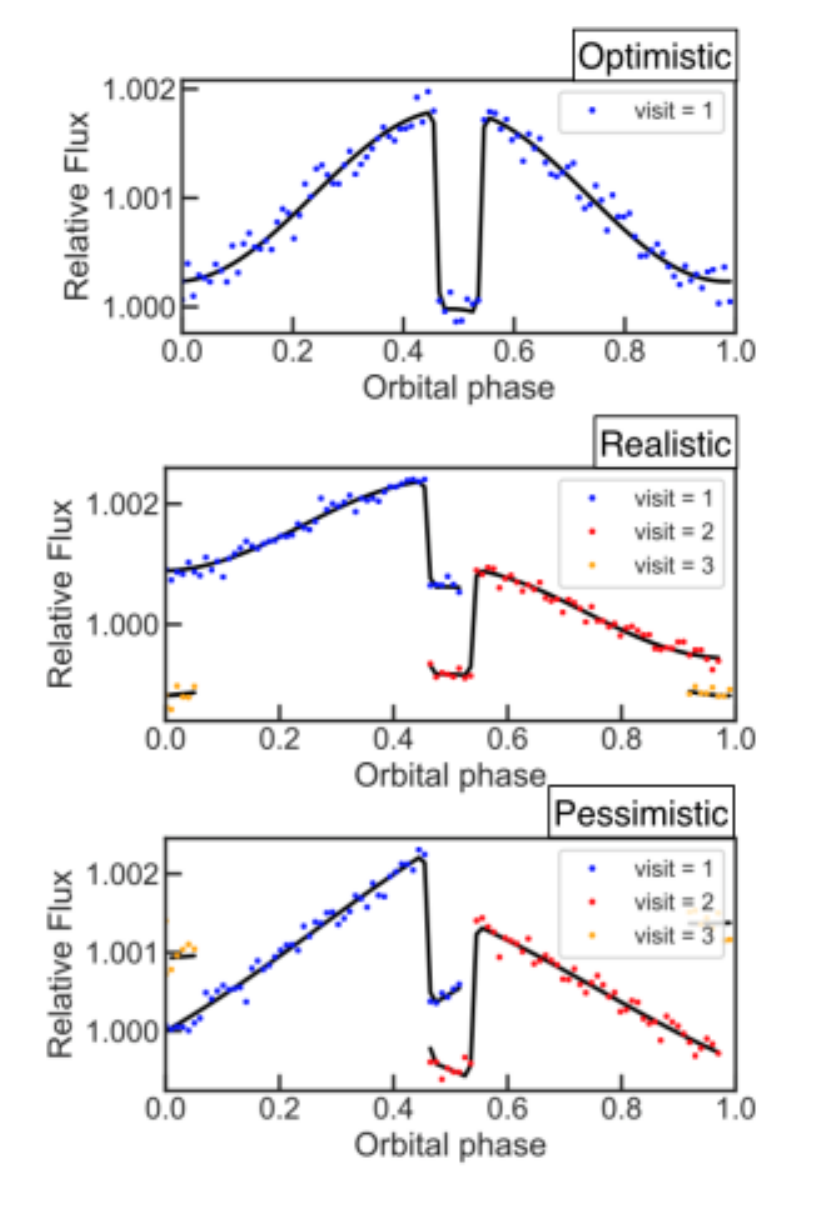}
\end{center}  
\caption{Simulated phase curves for WASP-43b for different assumptions about instrument performance. The optimistic case is a continuous stare with no instrument systematics, the realistic case is a phase curve split into  10-hour stares with 10$\%$ overlap in phase coverage with randomly drawn offsets between each visit, and the pessimistic case adds a randomly drawn linear trend to each 10-hour visit.}
\label{figure_model2}
\end{figure}

For each case (optimistic, realistic, pessimistic), the simulated phase curves for WASP-43b with clouds based on the \texttt{THOR} results are shown in Fig. \ref{figure_model2}. We generated 500 instances of each phase curve with randomly generated photon shot noise calculated with \texttt{Ariel-Rad}\cite{mugnai20} and the assumed instrument systematic noise model. We calculated the broadband phase curve from AIRS Channel 0 (1.95 - 3.9 microns). The time resolution of the simulated light curves was 0.01 times the orbital period of the planet. 

We fit a single sinusoid as our nominal phase variation model. We also performed a fit with a double sinusoid model that also included the first harmonic; however, the inferred amplitude of the harmonic was consistent with zero, so we focus here on results from the single sinusoid case. 

We show the inferred sinusoid amplitudes and offsets for each case in Table \ref{table3}. For case (1), the optimistic phase curve with continuous phase coverage and no instrument systematics, we detected the phase curve amplitude at high confidence (52 sigma for WASP-43b) and measured the phase offset to an accuracy of 2-3 degrees. This precision is more than sufficient to achieve the science goals for phase curve observations. For example, the measured phase curve will constrain the atmospheric metallicity to better than 0.5 dex and enable a search for variability due to weather.

\begin{table}[!h] 
\begin{tabular}{|l|l|l|}
\hline 
WASP-43b & Amplitude (ppm) & Hotspot offset (degrees)\\ \hline
1. Optimistic & 1560 +/- 30 & 4.2 +/- 1.0 \\
2a. Realistic (10$\%$ overlap) & 1560 +/- 30 & 4.3 +/- 2.1 \\
2b. Realistic (25$\%$ overlap)& 1560 +/- 30 & 3.9 +/- 1.8 \\
3a. Pessimistic (10$\%$ overlap) & 990 +/- 250 & 5.9 +/- 3.5 \\
3b. Pessimistic (25$\%$ overlap) & 1520 +/- 150 & 4.6 +/- 3.2 \\   
\hline
\end{tabular}
\caption{Accuracy of retrieved phase curve amplitude and hotspot offset for different scenarios of systematic noise.}
\label{table3}
\end{table}

For case (2), the light curve with three visits with constant offsets between them, we find that we can recover the phase curve amplitude almost as well as for case (1). The estimated sinusoid amplitudes are consistent, and the precision on the hotspot offset increases by a factor of two.

The more pessimistic case (3) gives consistent results with the other two cases; however, the precision is noticeably worse. The phase curve amplitude is detected at 10 sigma confidence, but only if the visits overlap by 25$\%$ in phase coverage. We conclude that depending on the phase curve amplitude and overlap in phase coverage, the pessimistic noise scenario may not allow us to achieve our science goals.

The ideal scenario is that Ariel observes phase curves continuously. If this is not possible, multi-epoch phase curves are a viable alternative. The optimal strategy depends on the instrument noise performance. If the detector is stable and has minimal trends in time (change in flux less than 1e-3 over 24 hours), the phase curve can be split into visits with just 10$\%$ overlap in phase coverage with no negative impact on science goals. By contrast, if linear systematic trends are present, at least 25$\%$ overlap in phase coverage is required to measure the phase curve amplitude and offset at sufficient precision, and it may only be possible for larger amplitude phase curves. The science will still be possible for the most pessimistic case, but the number of observable phase curves will decrease due to the large overlap in phase coverage that is required.

Ariel's instrument systematics should be carefully assessed early on in the mission to determine the optimal strategy for phase curve observations.

\subsection{Retrieval of spectroscopic phase curves}

To better understand the capability of Ariel to reveal the global climate behaviour of planets, we performed 1D spectral retrievals on the disc average flux of WASP-43b produced by \texttt{THOR}. We chose to perform retrievals with a phase angle step of 60$^{\circ}$. The error on the observations was estimated using \texttt{Ariel-Rad} for an integration time of 1 hour. 

We used the radiative transfer and retrieval framework \texttt{NEMESIS} \cite{irwin08,Taylor2020} to perform 1D spectral retrievals. We use the correlated-k approach to model our spectra \cite{lacis91,Chubb2020}, which has been shown to be effective and accurate when compared to using the line-by-line or cross-section approaches \cite{garland19}. The line lists used in the retrieval are those used to calculate \texttt{THOR}'s phase curves, however, \texttt{NEMESIS} uses a more recently developed line list for H$_2$O \cite{Polyansky18}.

\begin{figure}[!h] 
\begin{center} 
	\includegraphics[width=12cm]{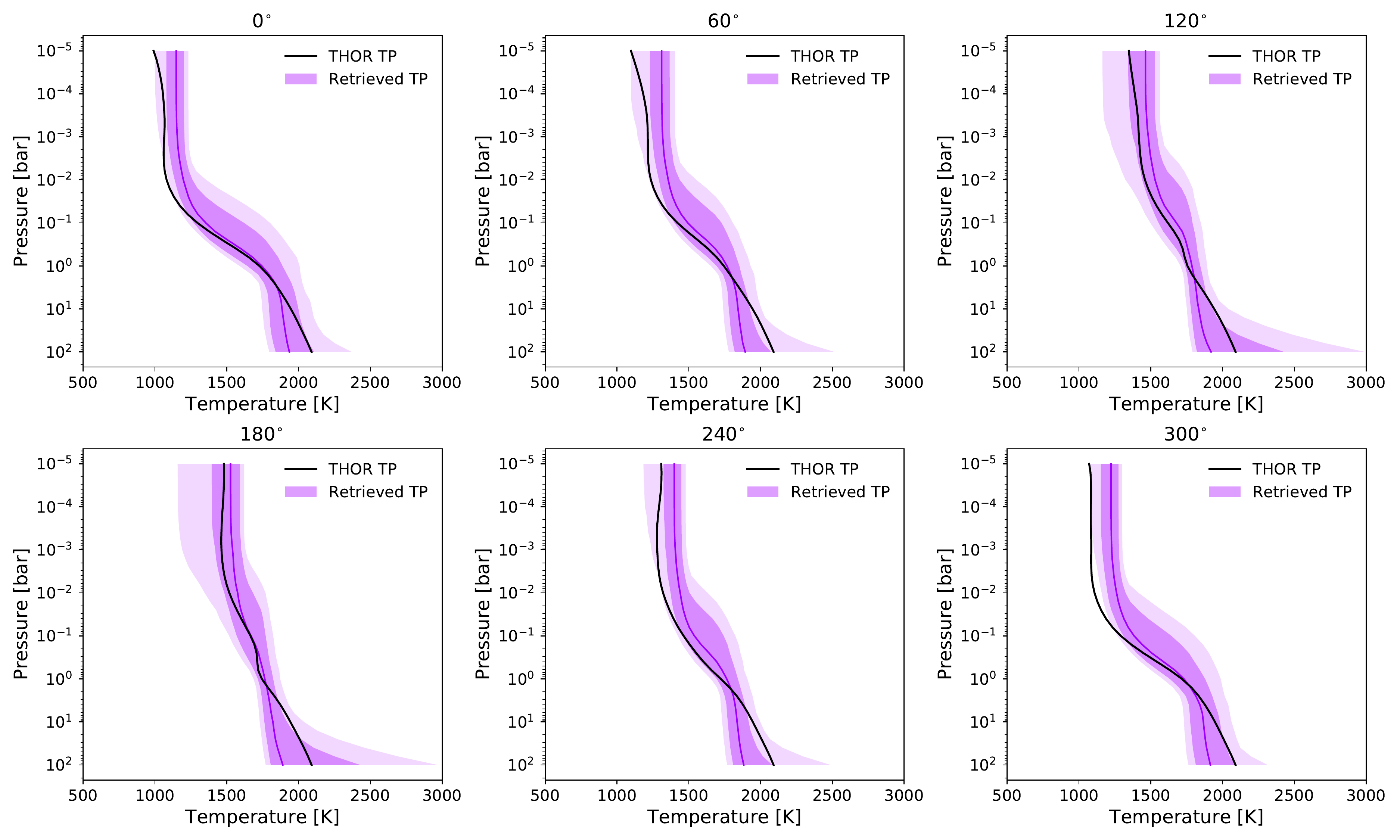}
\end{center}  
\caption{Retrieved temperature structures for the cloud free WASP-43b numerical experiment. We present the median, 1-sigma and 2-sigma intervals in decreasing opacity of purple, and the temperature structure from \texttt{THOR} is presented in black.}
\label{figure_model3}
\end{figure}

We model the atmosphere with a uniformly mixed vertical chemical structure and a temperature profile following the description of \cite{guillot10} and \cite{line13}. Fig. \ref{figure_model3} shows the retrieved temperature profiles for the cloud-free WASP-43b simulation. The temperature retrieval shows that we can generally retrieve the gradient of the profile within one sigma and the vertical shape of the profile within two sigma, apart from at a phase angle of 300$^{\circ}$ which is slightly outside of 2 sigma at the lowest pressures. 

\begin{figure}[!h] 
\begin{center} 
	\includegraphics[width=9cm]{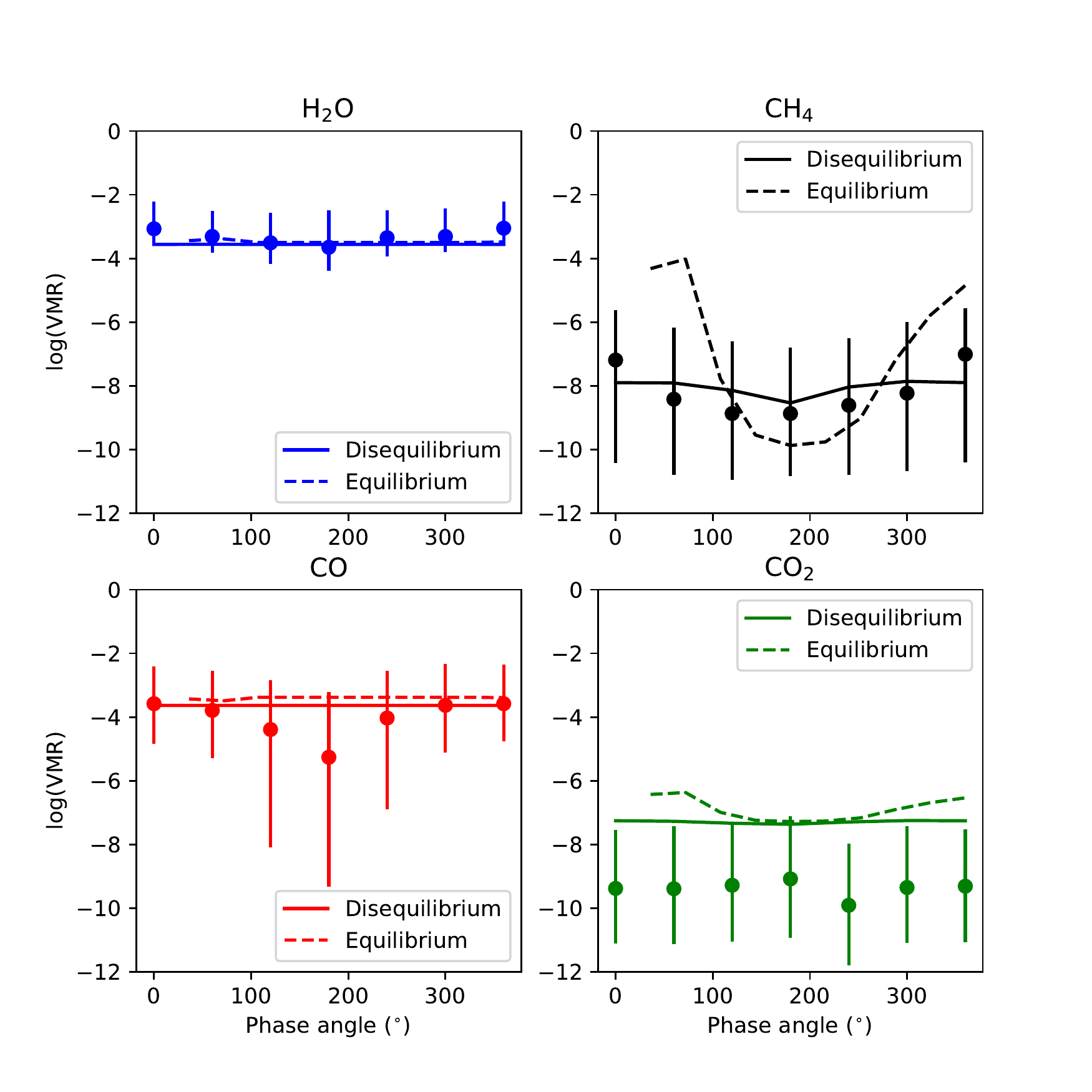}
\end{center}  
\caption{Retrieved abundances (VMR) and associated uncertainty for the cloud-free WASP-43b simulations. They are shown by the vertical lines for each molecule. The horizontal solid and dashed lines are the disequilibrium and equilibrium abundances respectively, the values are taken to be around $\sim$0.1 to 0.01 bar, the region in which our simulations are sensitive to. Phase 0$^\circ$ and phase 180$^\circ$ correspond to the transit and the eclipse respectively. Synthetic observations use disequilibrium chemistry.}
\label{figure_model4}
\end{figure}

In this retrieval exercise, we also test if we can distinguish between atmospheres in equilibrium and disequilibrium. The disequilibrium profiles are obtained from \texttt{THOR} simulations with solar metallicity and C/O. As seen in Fig. \ref{figure_model4}, we find that our retrieved chemical abundances are within one sigma of the input value, with the tightest constraints being on the water abundance. 
The water abundance in our simulation is 3.6$\times 10^{-4}$, compatible with HST constraints of $0.32-16\times10^{-4}$ \cite{kreidberg14b}.
We show that for H$_2$O and CO it is not possible to distinguish between the disequilibrium and equilibrium case, but it is possible for CO$_2$ and CH$_4$, with the greatest difference being for CH$_4$. It may however only be possible to put an upper limit on CH$_4$ and CO$_2$ due to their low abundance which results in less of an impact on the observed spectrum. H$_2$O abundance is retrieved with a precision of 3 dex for 1 hour observations. For 3 hour observations (6 different phases for WASP-43b), H$_2$O abundance is retrieved with a precision of 3 dex. The global mean H$_2$O abundance is retrieved with a precision of around 60$\%$ over the full orbit. This precision is sufficient to address the SQ2 for spectroscopic phase curves. We see from the chemistry retrieval that the error bars for CO are large for a phase angle of 180 degrees, this is due to the isothermal nature of the temperature-pressure profile in the region of the atmosphere to which we are sensitive to (i.e., around 0.1 bar), hence the features would appear muted and have a greater uncertainty.

\begin{figure}[!h] 
\begin{center} 
	\includegraphics[width=9cm]{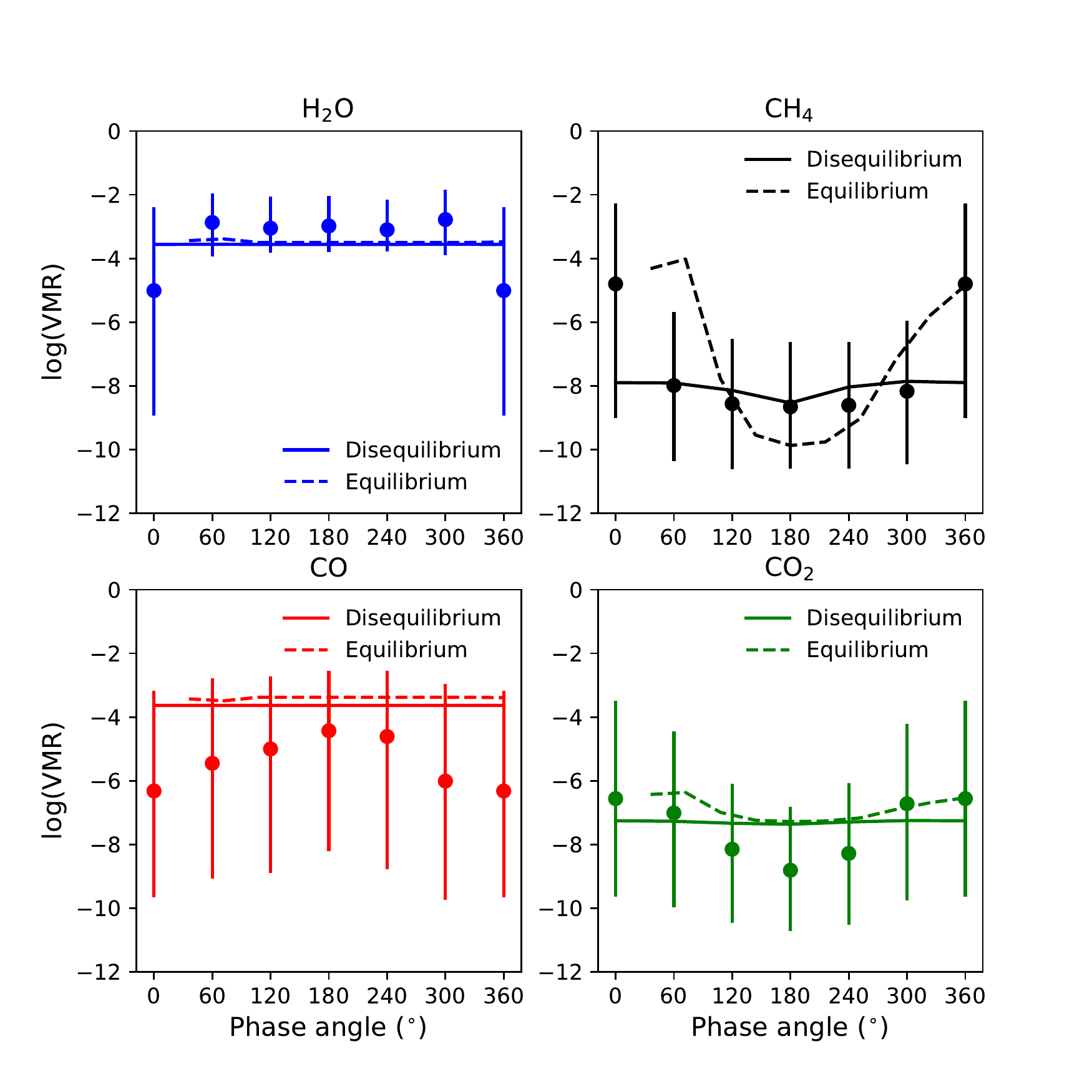}
\end{center}  
\caption{Retrieved abundances (VMR) and associated uncertainty for the cloudy WASP-43b simulations. They are shown by the vertical lines for each molecule. The horizontal solid and dashed lines are the disequilibrium and equilibrium abundances respectively, the values are taken to be around $\sim$0.1 to 0.01 bar, the region in which our simulations are sensitive to. Phase 0$^\circ$ and phase 180$^\circ$ correspond to the transit and the eclipse respectively. Synthetic observations use disequilibrium chemistry.}
\label{figure_model6}
\end{figure}

\begin{figure}[!h] 
\begin{center} 
	\includegraphics[width=12cm]{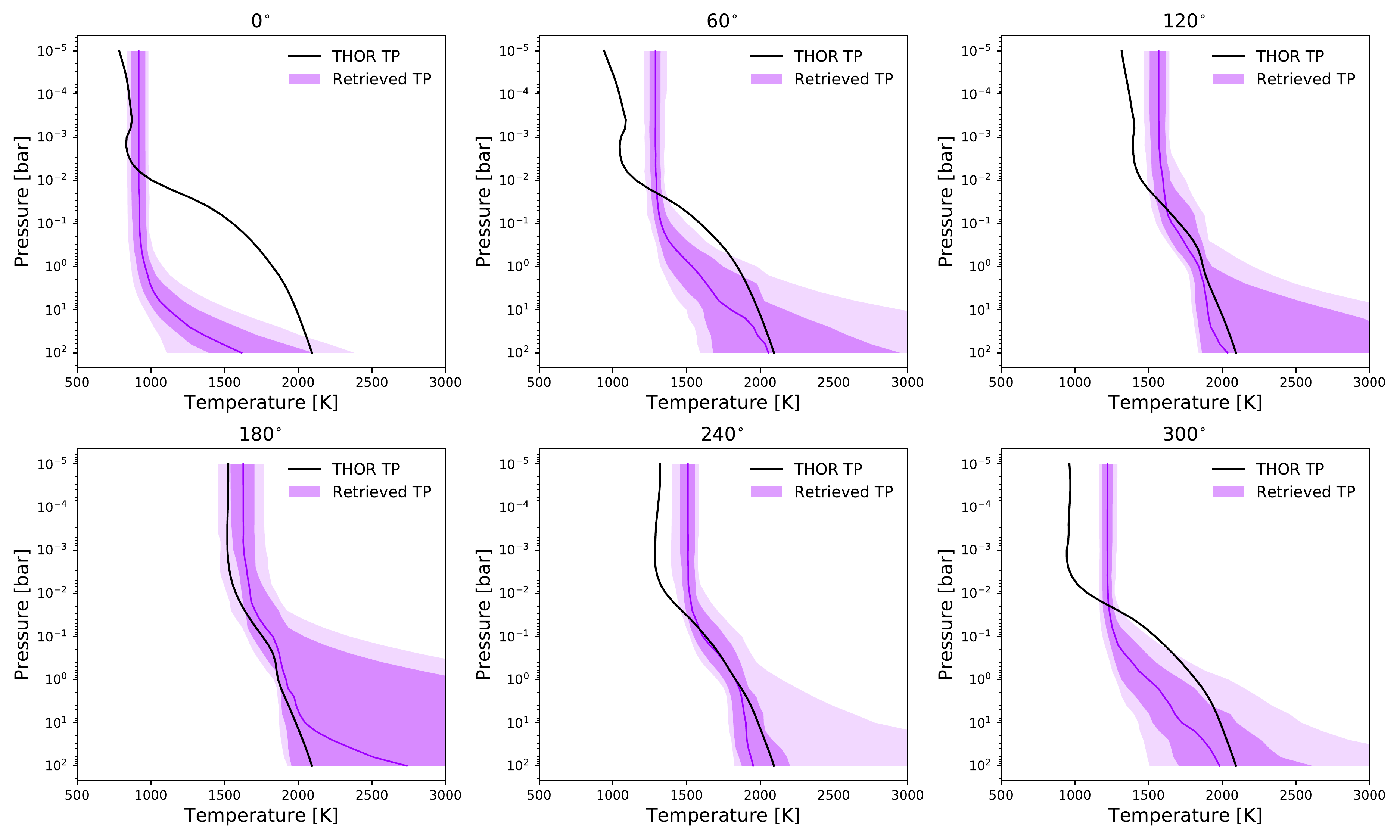}
\end{center}  
\caption{Retrieved temperature structures for the cloudy WASP-43b numerical experiment. We present the median, 1-sigma and 2-sigma intervals in decreasing opacity of purple, and the temperature structure from \texttt{THOR} is presented in black.}
\label{figure_model5}
\end{figure}

We perform the same retrieval exercise for an atmosphere that is cloudy. The 3D cloudy atmosphere was built from the simulations in \cite{mendonca18a} and shown in Fig. \ref{figure_model1}. To model the cloud in our retrieval framework we consider 3 extra parameters: the opacity of the cloud, the fractional scale height, and the cloud base pressure. We also assume that the cloud is non-scattering and grey in nature. We recognise that considering the scattering properties of the cloud could help to improve the retrieved atmospheric properties \cite{Taylor2020b}. The retrieved abundances in Fig. \ref{figure_model6} demonstrate that we are able to detect and constrain the abundance of H$_2$O with comparable precision to the cloud free scenario. 
We find that the other molecules are trickier to constrain due to the muted features introduced by the cloud. The retrieved thermal profiles in Fig. \ref{figure_model5} demonstrate the inability to recover the shape of the thermal structure for the cloudy atmosphere. The shape of the thermal structure calculated using \texttt{THOR} is not able to be captured by a Guillot-style parameterisation, given that the former is a state-of-the-art thermal structure parameterisations used in exoplanet emission retrievals. In other words, we demonstrate that observations from future instruments will require a more complex vertical temperature parameterisation for cloudy atmospheres.

\section{Summary and conclusions}
\label{sec:6}
Phase curves of thermal emission and reflected light are a powerful technic to characterize exoplanetary atmospheres. They provide insights into the atmospheric dynamics and longitudinal variations of the thermal structure, the chemical composition and clouds. In the next decade, JWST will likely observe a dozen planets with phase curves, preferentially smaller and cooler planets than Ariel can efficiently observe. 

We defined 5 science questions: 1) heat redistribution, 2)  longitudinal composition $\&$ temperature variations, 3) atmospheric composition of low-mass planets, 4) albedo $\&$ clouds and 5) time variability, to investigate with phase curves. They are focused on the coupling between atmospheric dynamics, chemical composition, thermal structure, and clouds. For most of these science questions, the requirement is to reach a SNR$>$10, i.e. detection at 10 sigma of photometric phase curves (at Tier 1 resolution), assuming no heat redistribution.

We constructed a list of 44 potential targets for phase curves from the Ariel general target list. These planets fulfill the requirements for the science questions and were divided into four size categories (rocky, sub-Neptunes, Neptunes and giants). Based on this potential target list, we expect that 35-40 exoplanets could be observed if 10$\%$ of Ariel science time were dedicated to phase curves.
According to this study, Neptune-size and giant planets are excellent targets for Ariel. They comprise $\sim$80$\%$ of planets in the target list. However, our observing plan only partially answers SQ4 (albedo $\&$ clouds), mostly based on reflected light. For giant planets (for which there is a multitude of good targets), an ideal strategy would be to choose those which will benefit from a precise visible phase curve observed by TESS, CHEOPS or PLATO. In particular, there would be a strong synergy by combining the hundreds of PLATO visible phase curves with Ariel infrared phase curves. 

The study of the effect of systematics shows that multi-epoch observations reduce the precision of phase-curve reconstruction. That performance degradation is drastic when a linear trend is present in the instrumental systematic effects.We therefore recommend performing continuous observations, starting before an eclipse and ending after the following eclipse. We also recommend to carefully assessing Ariel’s instrument systematic effects early on in the mission.Depending on their nature, we could consider multi-epoch phase curves, filling gaps in the observing scheduling.

The atmospheric retrieval of a simulated spectroscopic phase curves of cloud-free WASP-43b allows to measure the abundance of water within 0.5 dex for a 3-h observation. Combining measurements over the whole orbit, the mean abundance is retrieved with a 60$\%$ precision. In addition, the spectroscopic phase curve of WASP-43b allows distinguishing chemical equilibrium and disequilibrium thanks to CH$_4$ and CO$_2$ variations. It is important to note that all the giant planets on our target list have a SNR for a full phase curve similar to or higher than WASP-43b. This would imply a precision for atmospheric retrieval generally better than our simulated case for WASP-43b.

To conclude, the Ariel mission is a unique opportunity to perform a statistical survey of exoplanet phase curves, mostly warm/hot gaseous planets. These observations will be complemented by Ariel transit/eclipse spectroscopy and by measurements from other telescopes (e.g. TESS, CHEOPS, PLATO, JWST and ELTs). Together, they will provide insights into the global climate of these planets as well as a context for the interpretation of all Ariel observations.

\begin{acknowledgements}
B.C. acknowledges financial support from CNES.
J.M.M. work on Ariel is supported by PRODEX grant (PEA: 4000127377).
T.J.B. acknowledges support from the McGill Space Institute Graduate Fellowship, the Natural Sciences and Engineering Research Council of Canada’s Postgraduate Scholarships-Doctoral Fellowship, and from the Fonds de recherche du Québec – Nature et technologies through the Centre de recherche en astrophysique du Québec.
L.V.M. and E.P. were funded by the ASI grant n. 2018.22.HH.O. 
C.A. Haswell’s work on Ariel is supported by STFC under grant ST/T00178X/1
\end{acknowledgements}

\section{Appendix: Expression of the SNR for phase curves}

\subsection{A simple metric for phase curves}
We assume that the relative thermal flux variation with time (for an orbital period $P$) is given by:

\begin{equation}
F(t)= \frac{F_{\rm max}}{F_{\rm star}}\times \frac{\left(\cos(2\pi t/P)+1 \right)}{2}
\end{equation}
Phase curve data can be fitted with a cosine function $f(t)=A\times \cos(2 \pi t/P) + B$.
For $P\gg$1h, the uncertainty on the amplitude $A$ is:
\begin{equation}
\delta A= \sqrt{\frac{2}{P}}\sigma_{\rm 1h}
\end{equation}
where  $\sigma_{\rm 1h}$ is the noise (in ppm) in the given spectral or photometric band for a 1-hour observation of the host star and obtained with Ariel-Rad, and $P$ is expressed in hours.
With $A=\frac{F_{\rm max}}{2F_{\rm star}}$, the SNR of the full phase curve is:
\begin{equation}
SNR_{\rm orbit} = \frac{A}{\delta A} = 0.5\times SNR_{\rm 1h}\times \sqrt{P/2}
\end{equation}
where $SNR_{\rm 1h}=\frac{F_{\rm max}}{F_{\rm star}}\frac{1}{\sigma_{\rm 1h}}$ is the SNR for a 1-hour observation at full phase given in equation (1).

\subsection{Comparison to Spitzer data of LHS 3844b}
We compared our expressions (4) and (5) to the analysis of the Spitzer phase curves of LHS 3844b \cite{kreidberg19}. In this study, the planet-to-star flux variation is binned over 25 equally spaced intervals over the orbital period with 1-sigma uncertainties of $\sim$50 ppm. The peak-to-trough amplitude of the phase variation is 350$\pm$40 ppm from MCMC fitting. This corresponds to a SNR of $\sim$9. Applying formula (4) and (5), we find an uncertainty of the peak-to-trough amplitude of $\sim$30 ppm and a SNR of 12. We note that our basic fitting tends to underestimate the uncertainty on the amplitude of the phase curve since it assumes a perfect sine wave and does not take into account the transit and the eclipse. In addition, our formula gives the statistical uncertainty, so there would naturally be some deviation compared to a given dataset. Finally,
the uncertainty on the peak-to-trough sine amplitude for real LHS 3844b data is higher than predicted by the idealized model because it was simultaneously fit with an instrument systematic noise model. The instrument systematics for Ariel are expected to be much less severe than for Spitzer, so the Ariel uncertainties are expected to match those calculated with Equation (4).
Using noise estimations from Ariel-Rad, we predict a SNR of $\sim$13 with Ariel for the same observing duration ($\sim$3.8 days) and the same spectral range (4-5 $\mu$m) as Spitzer. The SNR is consistent with our previous estimation and it would rises up to 17 with AIRS-CH1. We conclude that our metric compares favourably with previous Spitzer observations and analysis of LHS 3844b.

%
%

\bibliographystyle{spphys}       

%
%

\end{document}